\documentclass[aps,prd,superscriptaddress,showpacs,reprint]{revtex4-1}
\usepackage[utf8]{inputenc}
\usepackage{textcomp}
\usepackage[dvipsnames]{xcolor}
\usepackage{amssymb,amsfonts,amsmath}
\usepackage[sort&compress]{natbib}
\usepackage[english]{babel}
\usepackage{braket}
\usepackage{graphicx}
\usepackage{bm}
\usepackage{bbm}
\usepackage{braket}
\usepackage{IEEEtrantools}
\usepackage{pstricks,pst-coil,pst-node,pst-circ,pst-plot,pst-3dplot,
pst-solides3d,pst-sigsys,pstricks-add,pst-eucl}
\allowdisplaybreaks[1]
\def\biggg#1{{\hbox{$\left#1\vbox to30.0pt{}\right.$}}}

\def\Biggg#1{{\hbox{$\left#1\vbox to50.0pt{}\right.$}}}

\DeclareMathOperator{\tr}{tr}

\begin{document}

\title{Low-energy limit of the $\bm{O(4)}$ quark-meson model\\
from the functional renormalization group approach}

\author{J\"urgen Eser}
\email[]{eser@th.physik.uni-frankfurt.de}
\affiliation{Institut f\"ur Theoretische Physik, 
Johann Wolfgang Goethe-Universit\"at, 
Max-von-Laue-Str.\ 1, 
D-60438 Frankfurt am Main, Germany}

\author{Florian Divotgey}
\email[]{fdivotgey@th.physik.uni-frankfurt.de}
\affiliation{Institut f\"ur Theoretische Physik, 
Johann Wolfgang Goethe-Universit\"at, 
Max-von-Laue-Str.\ 1, 
D-60438 Frankfurt am Main, Germany}

\author{Mario Mitter}
\email[]{mitter@bnl.gov}
\affiliation{Department of Physics, Brookhaven National Laboratory, Upton, NY 11973, USA}

\author{Dirk H.\ Rischke}
\email[]{drischke@th.physik.uni-frankfurt.de}
\affiliation{Institut f\"ur Theoretische Physik, 
Johann Wolfgang Goethe-Universit\"at, 
Max-von-Laue-Str.\ 1, 
D-60438 Frankfurt am Main, Germany}
\affiliation{Interdisciplinary Center for 
Theoretical Study and Department of
Modern Physics, 
University of Science and Technology of China, Hefei,
Anhui 230026, China}

\date{\today}

\begin{abstract}
We compute the low-energy limit of the $O(4)$-symmetric 
quark-meson model as an effective field theory for Quantum 
Chromodynamics (QCD) within the Functional Renormalization Group (FRG) 
approach. In particular, we analyze the renormalization
group flow of momentum-dependent pion self-interactions beyond the local
potential approximation. The numerical results for these couplings 
obtained from the FRG are confronted with a recent tree-level study. 
Additionally, their effect on the wave-function renormalization factors and 
the curvature masses is investigated.
\end{abstract}

\maketitle

\section{Introduction}
\label{sec:introduction}

The fundamental dynamics of the strong interaction is described by
QCD. In the case of $N_{f}$ massless quark flavors, the classical Lagrangian of QCD possesses a 
global $SU(N_{f})_{L} \times SU(N_{f})_{R} \times U(1)_{L} \times U(1)_{R}$ flavor
symmetry.
Due to an anomaly \cite{tHooft:1986ooh}, the $U(1)_{L} \times U(1)_{R}$
part of the symmetry is broken to $U(1)_{V} \equiv U(1)_{L+R}$, which corresponds 
to quark number conservation. We exclude this $U(1)_{V}$ symmetry 
from the following discussion, since it is trivially fulfilled in models with 
hadronic degrees of freedom, like the quark-meson model. 
For this work, the relevant flavor symmetry is then 
given by $SU(N_{f})_{L} \times SU(N_{f})_{R}$, the so-called chiral symmetry, 
which is further broken both explicitly and spontaneously.

The experimentally observed hadrons can be grouped 
into the irreducible representations of $SU(N_{f})_{V}\equiv SU(N_{f})_{L+R}$, 
but not into those of $SU(N_{f})_{L}\times SU(N_{f})_{R}$. This signals
the spontaneous breakdown of chiral symmetry 
to its diagonal flavor subgroup $SU(N_{f})_{V}$. 
An immediate consequence of this symmetry-breaking 
mechanism is the occurrence of $N_{f}^{2}-1$ 
(pseudo-)Nambu-Goldstone bosons (pNGBs). The fact that 
these bosons are only very light instead
of being massless indicates that the physical quarks have 
small finite masses, which break
chiral symmetry explicitly. Throughout the rest of this work, 
we restrict ourselves to the 
case of two dynamical quark flavors, $N_{f} = 2$, where the resulting
three pNGBs are identified with the pion isotriplet $\vec{\pi}$.

An important property of QCD is that its coupling $\alpha_{S}$ becomes 
large at low energies. This implies that perturbation theory cannot 
be used to study the low-energy regime of this theory. Therefore, one
has to use methods that do not rely on a perturbative expansion in 
powers of $\alpha_{S}$. One such possibility is given by Effective Field Theories (EFTs). A 
crucial guiding principle in the construction of 
EFTs for QCD is chiral symmetry, which 
can either be realized in a linear\ \cite{Schwinger:1957em, GellMann:1960np, 
Gasiorowicz:1969kn} or a nonlinear way\ \cite{Weinberg:1968de, Coleman:1969sm, 
Callan:1969sn}. The latter way of realizing a symmetry results in an EFT 
describing the interaction of pNGB fields, i.e., of pions for $N_{f}=2$, 
among themselves. There, the pNGBs explicitly enter the theory as local coordinates 
parametrizing the vacuum manifold of the theory. 

The most prominent EFT is given by Chiral Perturbation
Theory (ChPT)\ \cite{Gasser:1983yg, Gasser:1984gg}, whose connection to 
the low-energy regime of QCD has been studied in great detail in 
Ref.\ \cite{Leutwyler:1993iq}. ChPT is defined by a Lagrangian that 
contains all chiral invariants that are obtained from a systematic 
expansion in powers of derivatives of the pion fields. 
The coupling constants that enter this expansion are usually referred to 
as low-energy constants (LECs). Because this expansion contains arbitrarily 
high powers of derivatives of the pion fields, 
the resulting Lagrangian is not perturbatively renormalizable. 
However, the aforementioned expansion can also 
be understood as a power series in $p/(4\pi f_{\pi})$, where $p$ denotes 
the momentum of the pion fields and $f_{\pi}$ the pion 
decay constant. This power series is expected to converge for small enough 
pion momenta, and all infinities can be absorbed order by order into 
the LECs.
  
As already mentioned before, it is also possible to realize chiral 
symmetry in a linear way. The resulting models are referred 
to as Linear Sigma Models (LSMs)\ \cite{Koch:1997ei, Pisarski:1994yp}, 
which incorporate the pNGBs as well as their chiral partners 
on the same footing. Various versions of these models, also
with vector and axial-vector mesons\ \cite{Ko:1994en, Urban:2001ru}, 
were subject to comprehensive studies over the last decades. 
The relation between ChPT and hadronic models based on a linear realization of
chiral symmetry was studied in Refs.\ \cite{Bessis:1972sn, Jungnickel:1997yu, Jendges:2006yk}.
Concerning the low-energy limit, it was shown that the LECs of the most simple 
version of the LSM, including only the pion fields and the scalar sigma field 
$\sigma$, do not assume the same values as for QCD\ \cite{Gasser:1983yg}.

Recently, the so-called extended Linear Sigma Model (eLSM) was developed. 
This model contains all ground-state quark-antiquark mesons with (pseudo)scalar 
and (axial\hbox{-})\-vector quantum numbers up to $2$ GeV in mass. 
The Lagrangian of the eLSM respects all symmetries of QCD and also 
reflects their possible breaking patterns. The eLSM was studied for 
$N_{f} = 2, 3, 4$ in Refs.\  \cite{Parganlija:2010fz, Janowski:2011gt, 
Parganlija:2012fy, Eshraim:2014eka}. Baryons were included as well and
studied in vacuum \cite{Gallas:2009qp, Gallas:2013ipa, Olbrich:2015gln} 
and at finite density\ \cite{Gallas:2011qp, Heinz:2013hza}. 
From the requirement of dilatation symmetry and 
incorporating only positive semi-definite powers of the dilaton field, 
it follows that the eLSM Lagrangian contains only a finite number of terms. This 
results in a finite number of coupling constants and parameters, which 
have been determined in a global fit to experimentally measured masses and 
decay widths. It turns out that the eLSM is in remarkably good agreement 
with experimental data\ \cite{Parganlija:2012fy}, such as meson masses
and decay rates in the scalar-isoscalar sector, the $\eta$-$\eta'$
mixing angle, as well as branching ratios for $a_{0}$.

The question whether the low-energy limit of the eLSM
is consistent with that of QCD has been addressed in a recent tree-level 
study\ \cite{Divotgey:2016pst}. It turned out that the tree-level values of the
LECs of the eLSM are in good overall agreement with the ChPT values. 
The obvious next step is then to 
extend this study in order to check whether loop contributions 
to the LECs are negligible or not. In this paper,
we perform a first step in this direction by computing the loop corrections to the LECs
of the $O(4)$ quark-meson model as a simplified 
version of the mesonic eLSM to which we add a Yukawa coupling of 
mesons to quark fields. In future work, we will extend this
towards a study of loop corrections to the LECs of the full eLSM.

A framework that naturally generalizes the method presented in 
Ref.\ \cite{Divotgey:2016pst} is given by the Functional 
Renormalization Group (FRG)\ \cite{Wetterich:1992yh}, which has
been employed to consider low-energy effective theories 
as well as QCD under various (thermodynamical) 
aspects, see, for instance, Refs.\ \cite{Jungnickel:1995fp, 
Jungnickel:1995dt, Jungnickel:1996fd, 
Tetradis:2003qa, Schaefer:2004en, Braun:2009si, Kamikado:2012bt, 
Fejos:2012rz, Grahl:2013pba, Tripolt:2013jra, Mitter:2013fxa, 
Herbst:2013ufa, Grahl:2014fna, Braun:2014ata, Mitter:2014wpa, 
Eser:2015pka, Rennecke:2016tkm, Almasi:2016zqf, 
Jung:2016yxl, Strodthoff:2016pxx, Cyrol:2017ewj}.
By applying the FRG technique in this work, we shed light on the 
low-energy limit of the $O(4)$ quark-meson model. More precisely, we compute loop corrections 
to the tree-level low-energy couplings of this model 
by means of including momentum-dependent meson vertices
into the renormalization-group flow. In a first approximation
(the validity of which will be checked) we will restrict ourselves to pion self-interactions.
Our study can be viewed as an
extension of the work \cite{Jendges:2006yk} where,
in the local potential approximation (LPA) including 
scale-dependent wave-function renormalization factors 
as well as a scale-dependent Yukawa coupling between mesons and quarks,
it was shown that the FRG approach is able to produce the correct chiral logarithms in
the expressions for the pion decay constant and the pion mass.

This paper is organized as follows: Sec.\ \ref{sec:methods} 
introduces the concepts and methods that are used throughout 
this work. In Sec.\ \ref{sec:LSM}, we briefly review 
the $O(4)$ LSM and summarize its tree-level low-energy limit. After 
recalling the basics of the FRG approach in Sec.\ \ref{sec:FRG}, 
we derive the low-energy effective pion action for the $O(4)$ 
quark-meson model in Sec.\ \ref{sec:effpion}. The corresponding FRG 
flow equations are presented in the Appendix. Finally, in 
Sec.\ \ref{sec:results}, we show a comparison of the 
FRG study and the tree-level estimate for the low-energy couplings. 
Our conclusions and an outlook for future investigations 
are given in Sec.\ \ref{sec:summary}.

\section{Methods}
\label{sec:methods}

\subsection{$\bm{O(4)}$ Linear Sigma Model}
\label{sec:LSM}

As mentioned in Sec.\ \ref{sec:introduction}, the eLSM 
is a hadronic model that comprises (pseudo)scalar and 
(axial\hbox{-})\-vector mesons. Assignments 
of these fields to physical resonances can be found in 
Refs.\ \cite{Parganlija:2012fy, Divotgey:2016pst}. 

In the $O(4)$ limit, the mesonic degrees of freedom are 
described by the following matrix:
\begin{equation}
	\Phi = \sigma t_{0} + i\vec{\pi}\cdot\vec{t},\label{eq:phielsm}
\end{equation}
where $t_{0}=\mathbbmss{1}_{2}/2$ and $\vec{t}=\vec{\tau}/2$. The vector
$\vec{\tau}$ denotes the Pauli matrices. The generators
are normalized such that $\tr\!\left(t_{a}t_{b}\right) = \delta_{ab}/2$,
$a,b = 0,\ldots,3$.

Left- and right-handed chiral transformations act linearly on
the fields (\ref{eq:phielsm}) according to 
\begin{equation}
	\Phi \overset{U(2)_{L}\times U(2)_{R}}{\longrightarrow}
	U_{L}\Phi U_{R}^{\dagger}.
\end{equation}
The most general globally chirally symmetric Lagrangian that contains operators 
of dimension (up to) four and reproduces the chiral symmetry breaking pattern
found in Nature is given by
\begin{IEEEeqnarray}{rCl}
	\mathcal{L}_{O(4)} & = & \tr\left\{\left(\partial^{\mu}\Phi\right)^{\dagger}
	\partial_{\mu}\Phi\right\} - m_{0}^{2}\tr\left\{\Phi^{\dagger}\Phi\right\} 
	\nonumber\\
	& & - \lambda \left(\tr\left\{\Phi^{\dagger}\Phi\right\}\right)^{2} 
	+ \tr\left\{H\left(\Phi^{\dagger} + \Phi\right)\right\} ,
	\label{eq:o4lagrangian}
\end{IEEEeqnarray}
where, assuming exact isospin symmetry, $H=h_{\mathrm{ESB}}t_{0}$ and $h_{\mathrm{ESB}}\sim m_{u}=m_{d}$.
The explicit breaking of chiral symmetry due to non-vanishing quark 
masses is modeled by the term
\begin{equation}
	\tr\left\{H\left(\Phi^{\dagger}+\Phi\right)\right\}
	=h_{\mathrm{ESB}}\sigma ,
\end{equation}
which tilts the potential into the $\sigma$-direction. Evaluating
the traces, Eq.\ (\ref{eq:o4lagrangian}) becomes
\begin{IEEEeqnarray}{rCl}
	\mathcal{L}_{O(4)} & = & \frac{1}{2}
	\left(\partial_{\mu}\sigma\right)^{2} 
	+ \frac{1}{2}\left(\partial_{\mu}\vec{\pi}\right)^{2}
	- \frac{m_{0}^{2}}{2}\left(\sigma^{2} + \vec{\pi}^{2}\right) 
	\nonumber\\ 
	& & - \frac{\lambda}{4}
	\left(\sigma^{2} + \vec{\pi}^{2}\right)^2 
	+ h_{\text{ESB}}\sigma .\label{eq:o4lagrangian2}
\end{IEEEeqnarray}

The spontaneous breaking of chiral symmetry is reflected in
a non-vanishing vacuum expectation value $\sigma_{0}$ 
of the $\sigma$ field. The physical excitations of this field, 
corresponding to the $\sigma$ meson, are described by performing 
a shift in the Lagrangian (\ref{eq:o4lagrangian2}),
\begin{equation}
	\sigma\rightarrow\sigma_{0} + \sigma .\label{eq:shift}
\end{equation}
From this, one obtains the tree-level masses of the different mesons 
from terms quadratic in the fields,
\begin{IEEEeqnarray}{rCl}
	m_{\pi}^{2} & = & -m_{0}^{2} + \lambda\sigma_{0}^{2},\\
	m_{\sigma}^{2} & = & -m_{0}^{2} + 3\lambda\sigma_{0}^{2}.
\end{IEEEeqnarray}

As presented in Ref.\ \cite{Divotgey:2016pst}, the low-energy
limit of the eLSM can be obtained by successively integrating 
out all fields heavier than the pion. It turned out that this 
calculation can be performed analytically, if one restricts oneself 
to tree level. In this way, the low-energy effective Lagrangian of 
the eLSM in the $O(4)$ limit, which assumes the same mathematical
structure as ChPT, can be written as
\begin{IEEEeqnarray}{rCl}
	\mathcal{L}_{O(4),\text{eff}} & = & \frac{1}{2}\left(\partial_{\mu}
	\vec{\pi}\right)^{2}-\frac{1}{2}m_{\pi}^{2}\vec{\pi}^{2}
	+C_{1,O(4)}\left(\vec{\pi}^{2}\right)^{2} 
	\nonumber\\
	& & + \, C_{2,O(4)}\left(\vec{\pi}\cdot\partial_{\mu}
	\vec{\pi}\right)^{2}
	+C_{3,O(4)}\left[\left(\partial_{\mu}\vec{\pi}\right)
	\cdot \partial^{\mu}\vec{\pi}\right]^{2}
	\nonumber\\
	& & + \, \mathcal{O}(\pi^{6},\partial^{6}).
	\label{eq:o4lagrangianeff}
\end{IEEEeqnarray}
Using the tree-level masses, the parameters $\lambda$ and $m_{0}$
can be eliminated from the expressions for the low-energy couplings 
of the $O(4)$ LSM,
\begin{IEEEeqnarray}{rCl}
	C_{1,O(4)} & = & \frac{(m_{\sigma}^{2}-m_{\pi}^{2})^2}{8m_{\sigma}^{2}\sigma_{0}^{2}} \left( 1 -
	\frac{m_\sigma^2}{m_{\sigma}^{2}-m_{\pi}^{2}} \right) ,
	\label{eq:c1o4} \\
	C_{2,O(4)} & = & \frac{(m_{\sigma}^{2} 
	- m_{\pi}^{2})^{2}}{2 m_{\sigma}^{4}\sigma_{0}^{2}}, 
	\label{eq:c2o4} \\
	C_{3,O(4)} & = & \frac{(m_{\sigma}^{2} 
	- m_{\pi}^{2})^{2}}{2 m_{\sigma}^{6}\sigma_{0}^{2}} .
	\label{eq:c3o4}
\end{IEEEeqnarray}
These expressions slightly differ from the ones quoted in
Ref.\ \cite{Divotgey:2016pst}, where the equation of motion for
the free pion field was used to derive the low-energy 
couplings. Note that, for further purpose, we have separated
the first term in parentheses in Eq.\ (\ref{eq:c1o4}), which arises from
integrating out the $\sigma$ field, from the second one, which arises from the four-pion
interaction in the tree-level potential.

\subsection{Functional Renormalization Group}
\label{sec:FRG}

The FRG is an implementation of the Wilsonian renormalization 
principle. Changing from one energy scale to another, the integration 
of quantum and statistical fluctuations is performed momentum shell 
by momentum shell. The renormalization procedure thereby connects 
the microscopic interactions at an ultraviolet (UV) cutoff scale 
$\Lambda$ with the macroscopic physics through a sequence of effective 
theories.

Specifically, the FRG formulates a quantum field theory in terms of 
a differential equation. It focuses on the scale evolution of the 
effective average action $\Gamma_{k}$, where $k$ denotes the infrared 
(IR) cutoff introduced to the theory by adding a regulator 
function $R_{k}$ that acts as a momentum-dependent mass. 
The $k$-dependent $\Gamma_{k}$ interpolates between the renormalized classical 
action $S = \Gamma_{k\rightarrow \Lambda}$ 
and the full quantum effective action $\Gamma = \Gamma_{k\rightarrow 0}$ 
in the IR limit, where all fluctuations are integrated out. The effective 
action $\Gamma$ is the generating functional of one-particle irreducible 
vertex functions, thus containing all information about the quantum
theory.

For the upcoming FRG analysis, we switch to Euclidean space-time with 
a finite volume $\mathcal{V}$, leading to a discrete momentum spectrum. 
Lorentz indices $\mu = 0,1,2,3$ appear as lower indices. Furthermore, 
space-time integrations are indicated by a short-hand notation,
\begin{equation}
	\int_{\mathcal{V}}\mathrm{d}^{4}x = \int_{x}.
\end{equation}
Finally, we take the limit $\mathcal{V}\rightarrow \infty$ in all 
calculations.

The scale dependence of the effective average action is dictated by the 
Wetterich equation\ \cite{Wetterich:1992yh},
\begin{IEEEeqnarray}{rCl}
	\partial_{k}\Gamma_{k} & = & \frac{1}{2}\tr\left[
	\partial_{k} R_{k}\left(
	\Gamma^{(2)}_{k} + R_{k}\right)^{-1}\right] \nonumber\\
	& = & \frac{1}{2} \! \!
	\vcenter{\hbox{
	\begin{pspicture}(1.5,2.0)
		\psarc[linewidth=0.02](0.75,1.0){0.6}{113}{67}
		\pscircle[linewidth=0.03](0.75,1.6){0.25}
		\psline[linewidth=0.03](0.75,1.6)(0.92,1.77)
		\psline[linewidth=0.03](0.75,1.6)(0.58,1.77)
		\psline[linewidth=0.03](0.75,1.6)(0.58,1.43)
		\psline[linewidth=0.03](0.75,1.6)(0.92,1.43)
	\end{pspicture}
	}} \! \! ,\quad\ \label{eq:Wetterich} 
\end{IEEEeqnarray}
where we used a graphical interpretation of the propagator 
and the regulator insertion $\partial_{k}R_{k}$,
\begin{equation}
	\left(\Gamma^{(2)}_{k} + R_{k}\right)^{-1} = 
	\! \! \! \vcenter{\hbox{
	\begin{pspicture}[showgrid=false](1.5,0.5)
		\psline[linewidth=0.02](0.1,0.25)(1.4,0.25)
	\end{pspicture}
	}} \! \! ,\quad
	\partial_{k}R_{k} = \! \! \!
	\vcenter{\hbox{
	\begin{pspicture}[showgrid=false](1.5,0.8)
		\psline[linewidth=0.02](0.1,0.4)(0.5,0.4)
		\psline[linewidth=0.02](1.0,0.4)(1.4,0.4)
		\pscircle[linewidth=0.03](0.75,0.4){0.25}
		\psline[linewidth=0.03](0.75,0.4)(0.92,0.57)
		\psline[linewidth=0.03](0.75,0.4)(0.58,0.57)
		\psline[linewidth=0.03](0.75,0.4)(0.58,0.23)
		\psline[linewidth=0.03](0.75,0.4)(0.92,0.23)
	\end{pspicture}
	}} \! \! . \label{eq:rep}
\end{equation}
The second functional derivative of $\Gamma_{k}$ with 
respect to the fields, $\Gamma^{(2)}_{k}$, and 
the regulator $R_{k}$ are matrix-valued in field and momentum space as well 
as in all internal spaces, such as Dirac, color, and flavor space. 
The propagator appearing in Eq.\ (\ref{eq:Wetterich}) is fully field-dependent and partially dressed, 
i.e., it contains all fluctuations with momenta approximately
larger than the RG scale $k$. This is achieved
by the regulator insertion, which
is typically peaked around (squared) momenta of order $k^2$.

An infinite tower of 
coupled differential equations arises from Eq.\ (\ref{eq:Wetterich}): The flow of the effective average action 
is coupled to its second functional derivative. The corresponding 
equation for $\Gamma_{k}^{(2)}$, in turn, involves the third and fourth 
derivatives and, in general, the scale evolution of $\Gamma_{k}^{(n)}$ is 
influenced by derivatives up to order $n+2$. To obtain a closed set of 
differential equations, it is therefore necessary to truncate this system. 
For further details on the FRG approach, we refer to Refs.\ \cite{Morris:1993qb, 
Bagnuls:2000ae, Berges:2000ew, Pawlowski:2005xe, Gies:2006wv, Schaefer:2006sr, 
Kopietz:2010zz, vonSmekal:2012vx}.

Based on the discussion of the $O(4)$ LSM in the previous section, we choose 
the following ansatz for the Euclidean effective average action:
\begin{IEEEeqnarray}{rCl}
	\Gamma_{k} & = & \int_{x} \bigg\lbrace\frac{Z_{k}}{2} 
	\left(\partial_{\mu}\varphi\right)
	 \cdot \partial_{\mu}\varphi 
	 + U_{k}\left(\rho\right) - h_{\mathrm{ESB}}\sigma 
	\nonumber\\
	& & \qquad + \frac{Y_{1,k}}{8} \left[\partial_{\mu}\left(
	\varphi \cdot \varphi\right)
	\right]^{2}
	+ \frac{Y_{2,k}}{8}\, \varphi^{2}\left(\partial_{\mu}\varphi\right)
	 \cdot \partial_{\mu}\varphi \nonumber\\
	& & \qquad - \frac{X_{1,k}}{8} 
	\left[\left(\partial_{\mu}\varphi\right)
	\cdot \partial_{\mu}\varphi\right]^{2}
	- \frac{X_{2,k}}{8} 
	\left[\left(\partial_{\mu}\varphi\right)
	\cdot \partial_{\nu}\varphi\right]^{2}
	 \nonumber\\
	& & \qquad - \frac{X_{3,k}}{8} \,
	\varphi \cdot \left(\partial_{\mu}\partial_{\mu}\varphi\right)
	\left(\partial_{\nu}\varphi\right) \cdot \partial_{\nu}\varphi
	\nonumber\\
	& & \qquad - \frac{X_{4,k}}{8} \,
	\varphi^{2} \left(\partial_{\mu}\partial_{\nu}\varphi\right)
	\cdot \partial_{\mu}\partial_{\nu}\varphi
	\nonumber\\
	& & \qquad - \frac{X_{5,k}}{8} 
	\left(\varphi \cdot \partial_{\mu}\partial_{\mu}\varphi\right)^{2}
	- \frac{X_{6,k}}{8} \,
	\varphi^{2}\left(\partial_{\mu}\partial_{\mu}\varphi\right)^{2}
	\nonumber\\
	& & \qquad +\, \bar{\psi}\left(Z_{k}^{\,\psi}
	\gamma_{\mu}\partial_{\mu}+ y\Phi_{5}\right)\psi\bigg\rbrace ,
	\label{eq:genansatz}
\end{IEEEeqnarray}
with
\begin{equation}
	\rho = \varphi \cdot \varphi = \sigma^{2}+\vec{\pi}^{2}, \qquad
	\Phi_{5} = \sigma t_{0} + i\gamma_{5}\vec{\pi} \cdot \vec{t}.
\end{equation}
Here, we have introduced the vector $\varphi = (\sigma, \vec{\pi})$ 
as well as the scale-dependent constants $Z_{k}$, $Y_{1,k}$, 
$Y_{2,k}$, $X_{1,k}, \ldots , X_{6,k}$, and $Z_{k}^{\,\psi}$. 
The factors $Z_{k}$ and $Z_{k}^{\,\psi}$ describe the 
wave-function renormalization factors for scalars and fermions, respectively. 
The effective potential $U_{k}$ 
is a function of the $O(4)$ invariant $\rho$. The Yukawa coupling $y$ is 
assumed to be RG-scale independent [it was shown in Ref.\ \cite{Jendges:2006yk}
that this is a valid approximation in order to produce the chiral logarithms
for the pion decay constant and the pion mass]. The RG-scale 
invariant parameter $h_{\mathrm{ESB}} \neq 0$ leads 
to explicit symmetry breaking due to nonzero quark masses.

By definition, the effective average action\ (\ref{eq:genansatz}) is a functional of 
the classical fields. In a very common abuse of notation we do not introduce new symbols
for the classical fields. The physical vacuum expectation value in the absence of external sources will 
be denoted as $\sigma_{0}$. We will use this convention 
throughout the rest of this work.

Within the ansatz\ (\ref{eq:genansatz}) we have chosen the
following basis structures to span the full space of terms of order 
$\mathcal{O}(\varphi^{4},\partial^{2})$
and $\mathcal{O}(\varphi^{4},\partial^{4})$, respectively:
\begin{IEEEeqnarray}{rCl}
	\left[\partial_{\mu}\left(\varphi \cdot \varphi\right)\right]^{2},
	& \quad & \varphi^{2}\left(\partial_{\mu}\varphi\right)
	\cdot \partial_{\mu}\varphi , \label{eq:basis2}\\
	\left[\left(\partial_{\mu}\varphi\right)
	\cdot \partial_{\mu}\varphi\right]^{2}, & \quad &
	\left[\left(\partial_{\mu}\varphi\right)
	\cdot \partial_{\nu}\varphi\right]^{2}, \nonumber\\
	\varphi \cdot \left(\partial_{\mu}\partial_{\mu}\varphi\right)
	\left(\partial_{\nu}\varphi\right) \cdot \partial_{\nu}\varphi,
	& \quad & \varphi^{2} \left(\partial_{\mu}\partial_{\nu}\varphi\right)
	\cdot \partial_{\mu}\partial_{\nu}\varphi, \nonumber\\
	\left(\varphi \cdot \partial_{\mu}\partial_{\mu}\varphi\right)^{2},
	& \quad & \varphi^{2}\left(\partial_{\mu}
	\partial_{\mu}\varphi\right)^{2}. \label{eq:basis4}
\end{IEEEeqnarray}

Moreover, we include quark fluctuations into 
the FRG flow, which are not included in 
the analysis in Ref.\ \cite{Divotgey:2016pst}. Thus we extend the $O(4)$ LSM
from Sec.\ \ref{sec:LSM} to the $O(4)$ quark-meson model. However, these 
fermionic fluctuations do not affect the tree-level low-energy couplings 
in Eqs.\ (\ref{eq:c1o4}) -- (\ref{eq:c3o4}).

From the general truncation (\ref{eq:genansatz}) we obtain the 
specific ansatz
\begin{IEEEeqnarray}{rCl}
	\Gamma_{k} & = & \int_{x}\bigg\lbrace\frac{Z_{k}^{\,\sigma}}{2}
	\left(\partial_{\mu}\sigma\right) \partial_{\mu}\sigma + 
	\frac{Z_{k}^{\,\pi}}{2}
	\left(\partial_{\mu}\vec{\pi}\right) \cdot 
	\partial_{\mu}\vec{\pi} \nonumber\\ 
	& & \qquad +\, U_{k}(\rho)-h_{\mathrm{ESB}}\sigma 
	\nonumber\\
	& & \qquad +\, C_{2,k}\left(\vec{\pi} \cdot 
	\partial_{\mu}\vec{\pi}\right)^{2}
	+ Z_{2,k}\, \vec{\pi}^{2}\left(\partial_{\mu}\vec{\pi}\right)
	\cdot \partial_{\mu}\vec{\pi}
	\nonumber\\
	& & \qquad -\, C_{3,k} 
	\left[\left(\partial_{\mu}\vec{\pi}\right)
	\cdot \partial_{\mu}\vec{\pi}\right]^{2}
	- C_{4,k} \left[\left(\partial_{\mu}\vec{\pi}\right)
	\cdot \partial_{\nu}\vec{\pi}\right]^{2}
	 \nonumber\\
	& & \qquad -\, C_{5,k} \,
	\vec{\pi} \cdot \left(\partial_{\mu}\partial_{\mu}\vec{\pi}\right)
	\left(\partial_{\nu}\vec{\pi}\right) \cdot \partial_{\nu}\vec{\pi}
	\nonumber\\
	& & \qquad -\, C_{6,k} \,
	\vec{\pi}^{2} \left(\partial_{\mu}\partial_{\nu}\vec{\pi}\right)
	\cdot \partial_{\mu}\partial_{\nu}\vec{\pi}
	\nonumber\\
	& & \qquad -\, C_{7,k} 
	\left(\vec{\pi} \cdot \partial_{\mu}\partial_{\mu}\vec{\pi}\right)^{2}
	- C_{8,k} \,
	\vec{\pi}^{2}\left(\partial_{\mu}\partial_{\mu}\vec{\pi}\right)^{2}
	\nonumber\\
	& & \qquad +\, \bar{\psi}\left(Z_{k}^{\,\psi}
	\gamma_{\mu}\partial_{\mu}+ y
	\Phi_{5}\right)\psi
	\bigg\rbrace.
	\label{eq:specansatz}
\end{IEEEeqnarray}
Here, we have made the assignments
\begin{IEEEeqnarray}{rCl}
	Z_{k}^{\,\pi} & = & Z_{k} + \frac{Y_{2,k}}{4} \sigma_{0}^{2},\quad
	Z_{k}^{\,\sigma} = Z_{k} + Y_{1,k} \sigma_{0}^{2} 
	+ \frac{Y_{2,k}}{4} \sigma_{0}^{2},\nonumber \\
	C_{1,k} & = & -\frac{\lambda_{k}}{4},\quad
	C_{2,k} = \frac{1}{2} Y_{1,k},\quad
	Z_{2,k} = \frac{1}{8} Y_{2,k},\nonumber \\
	C_{3,k} & = & \frac{1}{8} X_{1,k},\quad
	C_{4,k} = \frac{1}{8} X_{2,k},\quad
	C_{5,k} = \frac{1}{8} X_{3,k},\nonumber\\
	C_{6,k} & = & \frac{1}{8} X_{4,k},\quad
	C_{7,k} = \frac{1}{8} X_{5,k},\quad
	C_{8,k} = \frac{1}{8} X_{6,k}. \label{eq:assign}
\end{IEEEeqnarray}

Note that we only extracted pure pion vertices from the higher 
couplings $Y_{1,k}$, $Y_{2,k}$, and $X_{1,k}, \ldots , X_{6,k}$. 
This is motivated by the fact that we want to keep track of 
the RG-scale evolution of exactly the same expressions that were 
produced in the tree-level approach, 
cf.\ Eq.\ (\ref{eq:o4lagrangianeff}). The terms $\sim Z_{2,k}$ and 
$\sim C_{4,k}, \ldots , C_{8,k}$ complete the structures 
from Eq.\ (\ref{eq:o4lagrangianeff}) to a full basis set, in
accordance with Eqs.\ (\ref{eq:basis2}) and (\ref{eq:basis4}).
This allows us to unambiguously project the momentum-dependent 
four-pion vertex onto these structures. 
All other terms proportional to $Y_{1,k}$, 
$Y_{2,k}$ or $X_{1,k}, \ldots , X_{6,k}$ 
(i.e., momentum-dependent sigma vertices and mixed sigma-pion
vertices) are neglected. This corresponds to an approximation
to the fully $O(4)$-symmetric ansatz\ (\ref{eq:genansatz})
based on the expectation that the nontrivial RG running of the
higher-derivative couplings will only set in roughly below the mass
threshold of the $\sigma$ field, i.e., that the latter will not significantly influence
the flow in the IR. Furthermore, the respective wave-function renormalization
factors for the $\sigma$ and the 
$\pi$ fields within this approximation, cf.\ the first line in Eq.\ (\ref{eq:assign}), 
will split as soon as $\sigma_{0}$ assumes a nonzero value. 
The momentum-independent four-pion 
interaction $C_{1,k} (\vec{\pi}^{2})^{2}$ is part of $U_{k}$ and, thus, 
not explicitly shown. We identify $C_{1,k}$ with a RG-scale
dependent version of the quartic coupling $\lambda$, cf. 
Eq.\ (\ref{eq:assign}).

Equation\ (\ref{eq:genansatz}) corresponds to a derivative expansion. Setting 
the factors $Z_{k}$ and $Z_{k}^{\,\psi}$ to one as well as 
$Y_{1,k}$, $Y_{2,k}$, and $X_{1,k}, \ldots , X_{6,k}$ to zero, 
we obtain its leading order, the LPA. In this case, the effective potential $U_{k}$ 
is the only $k$-dependent term. Going one step further, 
by taking the running of the wave-function renormalization factors $Z_{k}$ 
and $Z_{k}^{\,\psi}$ into account, the truncation is called LPA'. A general field 
and/or momentum dependence of $Z_{k}$ and $Z_{k}^{\,\psi}$ is suppressed. 
Hence, Eq.\ (\ref{eq:specansatz}) can be understood as a combination of the LPA' 
with the higher couplings $C_{2,k}$, $Z_{2,k}$, and $C_{3,k}, \ldots , C_{8,k}$.

We define the renormalized fields and couplings as
\begin{IEEEeqnarray}{rCl}
	\tilde{\sigma} & = & \sqrt{Z_{k}^{\,\sigma}} \sigma, \quad
	\tilde{\vec{\pi}} = \sqrt{Z_{k}^{\,\pi}} \vec{\pi}, \nonumber\\
	\tilde{\psi} & = & \sqrt{Z_{k}^{\,\psi}} \psi, \quad
	\tilde{\bar{\psi}} = \sqrt{Z_{k}^{\,\psi}} \bar{\psi}, \quad
	\tilde{h}_{\mathrm{ESB}} = \frac{h_{\mathrm{ESB}}}{\sqrt{Z_{k}^{\,\sigma}}}, \nonumber\\
	\tilde{C}_{1,k} & = & \frac{C_{1,k}}{({Z_{k}^{\,\pi}})^{2}}, \quad
	\tilde{C}_{2,k} = \frac{C_{2,k}}{({Z_{k}^{\,\pi}})^{2}}, \quad
	\tilde{Z}_{2,k} = \frac{Z_{2,k}}{({Z_{k}^{\,\pi}})^{2}}, \nonumber\\
	\tilde{C}_{3,k} & = & \frac{C_{3,k}}{({Z_{k}^{\,\pi}})^{2}}, \quad
	\tilde{C}_{4,k} = \frac{C_{4,k}}{({Z_{k}^{\,\pi}})^{2}}, \quad
	\tilde{C}_{5,k} = \frac{C_{5,k}}{({Z_{k}^{\,\pi}})^{2}},\nonumber\\
    \tilde{C}_{6,k} & = & \frac{C_{6,k}}{({Z_{k}^{\,\pi}})^{2}}, \quad
	\tilde{C}_{7,k} = \frac{C_{7,k}}{({Z_{k}^{\,\pi}})^{2}}, \quad
	\tilde{C}_{8,k} = \frac{C_{8,k}}{({Z_{k}^{\,\pi}})^{2}}, \nonumber\\
	\tilde{y}^{\sigma} & = & \frac{y}{Z_{k}^{\,\psi} \sqrt{Z_{k}^{\,\sigma}}},\quad
	\tilde{y}^{\pi} = \frac{y}{Z_{k}^{\,\psi} \sqrt{Z_{k}^{\,\pi}}}.
\end{IEEEeqnarray}
With these definitions, Eq.\ (\ref{eq:specansatz}) can be 
written as
\begin{IEEEeqnarray}{rCl}
	\Gamma_{k} & = & \int_{x}\bigg\lbrace\frac{1}{2}
	\left(\partial_{\mu}\tilde{\sigma}\right) \partial_{\mu}\tilde{\sigma} + 
	\frac{1}{2}
	\left(\partial_{\mu}\tilde{\vec{\pi}}\right) \cdot 
	\partial_{\mu}\tilde{\vec{\pi}} \nonumber\\ 
	& & \qquad +\, \tilde{U}_{k} - \tilde{h}_{\mathrm{ESB}}\tilde{\sigma} 
	\nonumber\\
	& & \qquad +\, \tilde{C}_{2,k}\left(\tilde{\vec{\pi}} \cdot 
	\partial_{\mu}\tilde{\vec{\pi}}\right)^{2}
	+ \tilde{Z}_{2,k}\, \tilde{\vec{\pi}}^{2}\left(\partial_{\mu}\tilde{\vec{\pi}}\right)
	\cdot \partial_{\mu}\tilde{\vec{\pi}}
	\nonumber\\
	& & \qquad -\, \tilde{C}_{3,k} 
	\left[\left(\partial_{\mu}\tilde{\vec{\pi}}\right)
	\cdot \partial_{\mu}\tilde{\vec{\pi}}\right]^{2}
	- \tilde{C}_{4,k} \left[\left(\partial_{\mu}\tilde{\vec{\pi}}\right)
	\cdot \partial_{\nu}\tilde{\vec{\pi}}\right]^{2}
	 \nonumber\\
	& & \qquad -\, \tilde{C}_{5,k} \,
	\tilde{\vec{\pi}} \cdot \left(\partial_{\mu}\partial_{\mu}\tilde{\vec{\pi}}\right)
	\left(\partial_{\nu}\tilde{\vec{\pi}}\right) \cdot \partial_{\nu}\tilde{\vec{\pi}}
	\nonumber\\
	& & \qquad -\, \tilde{C}_{6,k} \,
	\tilde{\vec{\pi}}^{2} \left(\partial_{\mu}\partial_{\nu}\tilde{\vec{\pi}}\right)
	\cdot \partial_{\mu}\partial_{\nu}\tilde{\vec{\pi}}
	\nonumber\\
	& & \qquad -\, \tilde{C}_{7,k} 
	\left(\tilde{\vec{\pi}} \cdot \partial_{\mu}\partial_{\mu}\tilde{\vec{\pi}}\right)^{2}
	- \tilde{C}_{8,k} \,
	\tilde{\vec{\pi}}^{2}\left(\partial_{\mu}\partial_{\mu}\tilde{\vec{\pi}}\right)^{2}
	\nonumber\\
	& & \qquad +\, \tilde{\bar{\psi}}\left(\gamma_{\mu}
	\partial_{\mu}+ 
	\tilde{y}^{\sigma}\tilde{\sigma} t_{0}
	+\, \tilde{y}^{\pi}i\gamma_{5}
	\tilde{\vec{\pi}} \cdot \vec{t} 
	\ \right)\tilde{\psi}\bigg\rbrace.
	\label{eq:renormansatz}
\end{IEEEeqnarray}
The fields and couplings in the effective potential also change 
accordingly, $U_{k}\rightarrow \tilde{U}_{k}$. In the IR limit, 
$k\rightarrow 0$, the fully renormalized expressions represent 
``measurable'' quantities. 

In this truncation, the partially conserved axial current (PCAC) relation is given by
\begin{equation}
	\mathcal{J}_{\mu,i}^{A} = 
	\sqrt{\frac{Z_{k}^{\,\pi}}{Z_{k}^{\,\sigma}}} \tilde{\sigma}_{0} \,
	\partial_{\mu} \tilde{\pi}_{i}
	+ \ldots = f_{\pi}\partial_{\mu}\tilde{\pi}_{i}, 
	\label{eq:PCAC}
\end{equation}
where $\mathcal{J}_{\mu,i}^{A}$ denotes the axial-vector current and 
\begin{equation}
	\tilde{\sigma}_{0} = \sqrt{Z_{k}^{\,\sigma}} \sigma_{0}
\end{equation}
denotes the renormalized vacuum expectation value. It is related to 
the pion decay constant via
\begin{equation}
	\tilde{\sigma}_{0} = f_{\pi}\sqrt{\frac{Z_{k}^{\,\sigma}}{Z_{k}^{\,\pi}}}. \label{eq:PCAC2}
\end{equation}
For a careful derivation of the PCAC 
relations in Eqs.\ (\ref{eq:PCAC}) and (\ref{eq:PCAC2}) see Appendix\ \ref{sec:PCAC}.

The flow equations for the scale evolution of the
truncation (\ref{eq:specansatz}) of the $O(4)$ quark-meson
model are rather lengthy and thus deferred to Appendix\ \ref{sec:floweqns}.

\subsection{Effective pion action}
\label{sec:effpion}

Following the strategy of Ref.\ \cite{Divotgey:2016pst},
we analytically integrate out the heavier $\tilde{\sigma}$ field 
to estimate the modified tree-level contribution to the low-energy 
couplings. This means that we have to reduce the effective action
in Eq.\ (\ref{eq:renormansatz}),
\begin{equation}
	\Gamma_{k} = \Gamma_{k}\left[\tilde{\sigma},\tilde{\vec{\pi}},
	\tilde{\bar{\psi}},\tilde{\psi}\right],
\end{equation}
to a theory solely consisting of pions, 
\begin{equation}
	\Gamma_{k} = \Gamma_{k}\left[\tilde{\vec{\pi}}\right].
\end{equation}

To this end, the quark fields are dropped in $\Gamma_{k}$ as
they do not influence the tree-level low-energy couplings. 
Afterwards, we eliminate the $\tilde{\sigma}$ field in the IR by exploiting
the quantum equation of motion,
\begin{equation}
	\frac{\delta\Gamma}{\delta\tilde{\sigma}} = 0.
	\label{eq:motion}
\end{equation}
Using the minimum condition
\begin{equation}
	\frac{\partial\tilde{U}_{k}}{\partial\tilde{\sigma}} = 
	\tilde{h}_{\mathrm{ESB}},
\end{equation}
the potential is assumed to take the form
\begin{IEEEeqnarray}{rCl}
	\tilde{U}_{k} & = & \frac{1}{2}M_{\sigma,k}^{2}\tilde{\sigma}^{2}
	+ \frac{1}{2}M_{\pi,k}^{2}\tilde{\vec{\pi}}^{2} 
	+ \frac{\tilde{\lambda}_{1,k}}{4}\tilde{\sigma}^{4}
	+ \tilde{\lambda}_{1,k}\tilde{\sigma}_{0}\tilde{\sigma}^{3} \nonumber\\
	& & + \frac{\tilde{\lambda}_{2,k}}{4}\left(\tilde{\vec{\pi}}^{2}\right)^{2}
	+ \frac{\tilde{\lambda}_{3,k}}{2}\tilde{\sigma}^{2}\tilde{\vec{\pi}}^{2}
	+ \tilde{\lambda}_{3,k}\tilde{\sigma}_{0}\tilde{\sigma}\tilde{\vec{\pi}}^{2},
	\label{eq:potential}
\end{IEEEeqnarray}
with
\begin{equation}
	\tilde{\lambda}_{1,k} = \frac{\lambda_{k}}{({Z_{k}^{\,\sigma}})^{2}},\quad
	\tilde{\lambda}_{2,k} = \frac{\lambda_{k}}{({Z_{k}^{\,\pi}})^{2}},\quad
	\tilde{\lambda}_{3,k} = \frac{\lambda_{k}}{Z_{k}^{\,\sigma}Z_{k}^{\,\pi}}.
\end{equation}
This assumption is motivated by the tree-level approach presented in 
Ref.\ \cite{Divotgey:2016pst}, where the authors restricted themselves to
a potential of the form (\ref{eq:o4lagrangian2}).
The renormalized masses are defined as
\begin{equation}
	M_{\sigma,k}^{2} = \frac{m_{\sigma,k}^{2}}
	{Z_{k}^{\,\sigma}} ,\quad
	M_{\pi,k}^{2} = \frac{m_{\pi,k}^{2}}
	{Z_{k}^{\,\pi}} ,\quad
	M_{\psi,k}^{2} = \frac{m_{\psi,k}^{2}}
	{({Z_{k}^{\,\psi}})^{2}} .
\end{equation}
At tree level, only the last interaction term in Eq.\ (\ref{eq:potential}) 
contributes to the equation of motion (\ref{eq:motion}) of the $\tilde{\sigma}$ 
field,
\begin{equation}
	\left(\partial_{\mu}\partial_{\mu} - M_{\sigma,k}^{2}\right)\tilde{\sigma} = 
	\tilde{\lambda}_{3,k}\tilde{\sigma}_{0}\tilde{\vec{\pi}}^{2}.
\end{equation}
Solving this equation for $\tilde{\sigma}$, we find
\begin{equation}
	\tilde{\sigma} = -\frac{\tilde{\lambda}_{3,k}\tilde{\sigma}_{0}}{M_{\sigma,k}^{2}}
	\left[1 + \frac{\partial_{\mu}\partial_{\mu}}{M_{\sigma,k}^{2}} + 
	\frac{\left(\partial_{\mu}\partial_{\mu}\right)^{2}}{M_{\sigma,k}^{4}} + 
	\mathcal{O}\left(\partial^{6}\right)\right]\tilde{\vec{\pi}}^{\,2}.
\end{equation}
Using this relation, we finally arrive at
\begin{IEEEeqnarray}{rCl}
	\Gamma_{k} & = &
	\int_{x}\bigg\lbrace\frac{1}{2}
	\left(\partial_{\mu}\tilde{\vec{\pi}}\right)
	\cdot \partial_{\mu}\tilde{\vec{\pi}} + 
	\frac{1}{2}M_{\pi,k}^{2}\tilde{\vec{\pi}}^{2} 
	- \tilde{C}^{\mathrm{total}}_{1,k}
	\left(\tilde{\vec{\pi}}^{2}\right)^{2}
	\nonumber\\
	& & \qquad +\, \tilde{C}^{\mathrm{total}}_{2,k}
	\left(\tilde{\vec{\pi}}\cdot
	\partial_{\mu}\tilde{\vec{\pi}}\right)^{2}
    + \tilde{Z}^{\mathrm{total}}_{2,k}\, \tilde{\vec{\pi}}^{2} 
	\left(\partial_{\mu}\tilde{\vec{\pi}}\right)\cdot
	\partial_{\mu}\tilde{\vec{\pi}} \nonumber\\
	& & \qquad -\, \tilde{C}^{\mathrm{total}}_{3,k}
	\left[\left(\partial_{\mu}\tilde{\vec{\pi}}\right) \cdot 
	\partial_{\mu}\tilde{\vec{\pi}}\right]^{2}
	\nonumber\\
	& & \qquad -\, \tilde{C}^{\mathrm{total}}_{4,k} 
	\left[\left(\partial_{\mu}\tilde{\vec{\pi}}\right)
	\cdot \partial_{\nu}\tilde{\vec{\pi}}\right]^{2}
	 \nonumber\\
	& & \qquad -\, \tilde{C}^{\mathrm{total}}_{5,k} \,
	\tilde{\vec{\pi}} \cdot 
	\left(\partial_{\mu}\partial_{\mu}\tilde{\vec{\pi}}\right)
	\left(\partial_{\nu}\tilde{\vec{\pi}}\right) \cdot 
	\partial_{\nu}\tilde{\vec{\pi}} \nonumber\\
	& & \qquad -\, \tilde{C}^{\mathrm{total}}_{6,k} \,
	\tilde{\vec{\pi}}^{2} \left(\partial_{\mu}
	\partial_{\nu}\tilde{\vec{\pi}}\right)
	\cdot \partial_{\mu}\partial_{\nu}\tilde{\vec{\pi}}
	\nonumber\\
	& & \qquad -\, \tilde{C}^{\mathrm{total}}_{7,k} 
	\left(\tilde{\vec{\pi}} \cdot \partial_{\mu}
	\partial_{\mu}\tilde{\vec{\pi}}\right)^{2}
	\nonumber\\
	& & \qquad -\, \tilde{C}^{\mathrm{total}}_{8,k} \,
	\tilde{\vec{\pi}}^{2}\left(\partial_{\mu}
	\partial_{\mu}\tilde{\vec{\pi}}\right)^{2}
	\bigg\rbrace .
\end{IEEEeqnarray}
The low-energy couplings $\tilde{C}^{\mathrm{total}}_{i,k}$ with
$i \in \lbrace 1,\ldots,8 \rbrace$ and
$\tilde{Z}^{\mathrm{total}}_{2,k}$ are given by
\begin{IEEEeqnarray}{rCl}
	\tilde{C}^{\mathrm{total}}_{i,k} & = & 
	\tilde{C}^{\mathrm{tree}}_{i,k} + \tilde{C}_{i,k},\\
	\tilde{Z}^{\mathrm{total}}_{2,k} & = & 
	\tilde{Z}^{\mathrm{tree}}_{2,k} + \tilde{Z}_{2,k},
\end{IEEEeqnarray}
where
\begin{IEEEeqnarray}{rCl}
	\tilde{C}^{\mathrm{tree}}_{1,k} & = & 
	\frac{(Z_{k}^{\,\sigma}M_{\sigma,k}^{2} 
	- Z_{k}^{\,\pi}M_{\pi,k}^{2})^{2}}
	{8({Z_{k}^{\,\pi}})^{2}M_{\sigma,k}^{2}\tilde{\sigma}_{0}^{2}}, 
	\label{eq:c1o4mod} \\
	\tilde{C}^{\mathrm{tree}}_{2,k} & = & 
	\frac{(Z_{k}^{\,\sigma}M_{\sigma,k}^{2} 
	- Z_{k}^{\,\pi}M_{\pi,k}^{2})^{2}}
	{2({Z_{k}^{\,\pi}})^{2}M_{\sigma,k}^{4}\tilde{\sigma}_{0}^{2}},
	\label{eq:c2o4mod} \\
	\tilde{C}^{\mathrm{tree}}_{3,k} & = & \frac{(Z_{k}^{\,\sigma}M_{\sigma,k}^{2} 
	- Z_{k}^{\,\pi}M_{\pi,k}^{2})^{2}}
	{2({Z_{k}^{\,\pi}})^{2}M_{\sigma,k}^{6}\tilde{\sigma}_{0}^{2}},
	\label{eq:c3o4mod} \\
	\tilde{C}^{\mathrm{tree}}_{5,k} & = & 
	\frac{(Z_{k}^{\,\sigma}M_{\sigma,k}^{2} 
	- Z_{k}^{\,\pi}M_{\pi,k}^{2})^{2}}
	{({Z_{k}^{\,\pi}})^{2}M_{\sigma,k}^{6}\tilde{\sigma}_{0}^{2}}, \\
	\tilde{C}^{\mathrm{tree}}_{7,k} & = & 
	\frac{(Z_{k}^{\,\sigma}M_{\sigma,k}^{2} 
	- Z_{k}^{\,\pi}M_{\pi,k}^{2})^{2}}
	{2({Z_{k}^{\,\pi}})^{2}M_{\sigma,k}^{6}\tilde{\sigma}_{0}^{2}}, \\
	\tilde{Z}^{\mathrm{tree}}_{2,k} & = & 
	\tilde{C}^{\mathrm{tree}}_{4,k} =
	\tilde{C}^{\mathrm{tree}}_{6,k} =
	\tilde{C}^{\mathrm{tree}}_{8,k} = 0. \label{eq:co4mod}
\end{IEEEeqnarray}
The superscript ``tree'' indicates that these loop-corrected contributions to
the low-energy couplings, apart from those explicitly written down
in the truncation (\ref{eq:specansatz}), are generated by the elimination of the
$\tilde{\sigma}$ field. As pointed out, this is done in close analogy 
to the tree-level calculation of Ref.\ \cite{Divotgey:2016pst}.

Neglecting the flow of the wave-function renormalization factors and the 
scale dependence of the meson masses during
the integration process,
\begin{IEEEeqnarray}{rCl}
	Z_{k}^{\,\sigma} = Z_{k}^{\,\pi} = 1, & \quad &
	\tilde{\sigma}_{0} \rightarrow \sigma_{0}, \nonumber\\ 
	M_{\sigma,k}^{2} \rightarrow m_{\sigma}^{2}, & \quad &
	M_{\pi,k}^{2} \rightarrow m_{\pi}^{2}, 
\end{IEEEeqnarray}
we reproduce the tree-level low-energy couplings of the $O(4)$ LSM
in Eqs.\ (\ref{eq:c1o4}) -- (\ref{eq:c3o4}). Note that, since
Eq.\ (\ref{eq:c1o4mod}) was obtained by integrating out the $\tilde{\sigma}$ field
at tree-level, we only obtain the
first term in parentheses in Eq.\ (\ref{eq:c1o4}), while the second term in that equation
corresponds to the fifth term on the right-hand side of Eq.\ (\ref{eq:potential}).

Instead of using the equation of motion for the $\tilde{\sigma}$ field
to derive the purely pionic theory, one could have arrived at the
same result by adding all diagrams that are one-particle reducible
with respect to $\tilde{\sigma}$-meson lines to the one-particle irreducible 
pion amplitudes. Obviously, this also applies to the previous result,
Eqs.\ (\ref{eq:c1o4}) -- (\ref{eq:c3o4}), as a special case.

\section{Results}
\label{sec:results}

The FRG flow equations for the effective potential, the wave-function 
renormalization factors, and the higher derivative couplings 
constitute a set of coupled partial differential equations. The effective
potential is numerically solved using a grid in $\sigma^{2}$, whereas the other
couplings are evaluated at the IR minimum $\sigma_{0,k_{\mathrm{IR}}}\equiv \sigma_{0}$.

We choose $\Lambda = 500$ MeV as UV cutoff. Recent investigations
\cite{Braun:2014ata, Mitter:2014wpa, Cyrol:2017ewj}, which are based on 
the dynamical hadronization technique 
\cite{Gies:2001nw, Gies:2002hq, Pawlowski:2005xe, Floerchinger:2009uf},
indicate that the actual range of validity of NJL-like models like the quark-meson
model is closer to $\Lambda = 300$ MeV. However, we also want to capture 
quark dynamics beyond the confinement scale 
$\Lambda_{\mathrm{QCD}} \simeq 200$ MeV, which is also the reasoning behind 
choosing cutoff scales as large as 1 GeV, see, e.g., 
Refs.\ \cite{Braun:2009si, Braun:2014ata, Mitter:2013fxa, Tripolt:2013jra}.

From a technical point of view, the inclusion of quarks is advantageous 
in the sense that, analogously to QCD, 
it allows us to start the FRG flow in an approximately symmetric 
regime. The fermionic fluctuations in Eq.\ (\ref{eq:Wetterich}) will 
drive $\sigma_{0,k}$ to larger nonzero values, 
similar to the NJL-model mechanism of chiral symmetry breaking,
which captures the low-energy dynamics of QCD
\cite{Braun:2014ata, Mitter:2014wpa, Cyrol:2017ewj}. 
The effective potential is then initialized as follows:
\begin{equation}
	U_{\Lambda}\left(\sigma^{2}\right) = \frac{m_{0,\Lambda}^{2}}{2}\sigma^{2} 
	+ \frac{\lambda_{\Lambda}}{4} \sigma^4 .
\end{equation} 
The parameters $m_{0,\Lambda}$ and $\lambda_{\Lambda}$ are
tuned such that the IR curvature masses and the pion decay constant 
are consistent with experimental data
\cite{Olive:2016xmw}. The concrete parameters $m_{0,\Lambda}$ and 
$\lambda_{\Lambda}$ that we used to produce the numerical results of 
this section are summarized in Tab.\ \ref{tab:UVpars}.
\begin{table}
	\caption{\label{tab:UVpars}UV parameters ($\Lambda = 500$ MeV).}
	\begin{ruledtabular}
		\begin{tabular}{lcccc}
		\textbf{Parameter} & $m_{0}$ & $\lambda$ & $h_{\mathrm{ESB}}$ & $y$ \\ 
		\colrule
		\textbf{Value} & $500$ MeV & $1.5$ & $2.2 \times 10^{6}\ \mathrm{MeV}^{3}$ & $9.0$
		\end{tabular}
	\end{ruledtabular}
\end{table}
The wave-function renormalization factors start at a value of 
one in the UV, while the higher couplings $C_{2,k}$, $Z_{2,k}$, and
$C_{3,k},\ldots , C_{8,k}$ are initialized as zero. 
They are only generated during the integration process.
From the scale-dependent minimum $\sigma_{0,k}$ of the effective 
potential we obtain the squared meson and quark masses,
\begin{IEEEeqnarray}{rCl}
	m_{\sigma,k}^{2} & = & 2 U_{k}'(\sigma_{0,k}^{2}) + 
	4\sigma_{0,k}^{2} U_{k}''(\sigma_{0,k}^{2}) ,\\
	m_{\pi,k}^{2} & = & 2 U_{k}'(\sigma_{0,k}^{2}) , \\
	m_{\psi,k}^{2} & = & \frac{y^{2}}{4}\sigma_{0,k}^{2} .
\end{IEEEeqnarray}

Already at this stage of the analysis we are able to make a general statement 
about the investigated theory: In the symmetric phase, meaning 
$\sigma_{0,k} = 0$, the masses of the $\sigma$ field and the pions are degenerate 
and the quark mass vanishes. In the presence of explicit symmetry
breaking, $h_{\mathrm{ESB}} \neq 0$, this can at best be achieved approximately
and $\sigma_{0,k}$ can only serve as an approximate order parameter for 
possible phase transitions.

\begin{figure}[ht]
	\centering
		\includegraphics{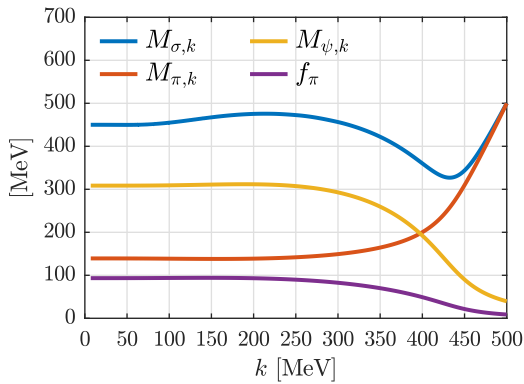}
	\caption{Scale evolution of the renormalized meson and quark masses 
	and the pion decay constant; $k_{\mathrm{IR}} = 7$ MeV.}
	\label{fig:masses}
\end{figure}

In Fig.\ \ref{fig:masses} we plot the renormalized masses 
and the vacuum expectation value as a function of the IR 
scale $k$.
As already stated in the last paragraph, $\sigma_{0,k}$ is close to zero 
for high energies ($k\rightarrow \Lambda$) and, consequently, so is 
the quark mass $M_{\psi,k}$. In the same region, the masses of the 
$\sigma$ and the pions are almost identical, as expected. With a mass 
of around 500 MeV they are effectively separated from the FRG flow, 
which is dominated by the light current quarks.

For a lower scale $k$, roughly between 400 and 450 MeV, the system undergoes a crossover transition
to the spontaneously broken phase, 
where the minimum of the effective potential is shifted towards 
higher values. At a scale of $k \simeq 460$ MeV, the splitting of 
the meson masses sets in and becomes increasingly stronger for  
decreasing $k$.

In the low-energy regime at $k = 150$ MeV and below, the pion and 
the quark masses as well as $\sigma_{0,k}$ settle at their IR 
values. For the $\sigma$ mass a scale
dependence persists until $k< 50$ MeV.

We stopped the FRG flow at $k = 7$ MeV, an arbitrarily chosen scale 
close to zero (approximately $1\%$ of the UV cutoff) at which 
all $k$-dependent 
quantities become constant. At this point, the pions 
have a mass of 139.3 MeV, the $\sigma$ field has a mass of 450.1 MeV, 
and the quarks acquire a constituent mass of 308.5 MeV. As required by 
the PCAC relations (\ref{eq:PCAC}) and (\ref{eq:PCAC2}), 
$\sigma_{0,k}\sqrt{Z_{k}^{\,\pi}}$ assumes a value of 93.4 MeV in the IR.

\begin{figure}[ht]
	\centering
		\includegraphics{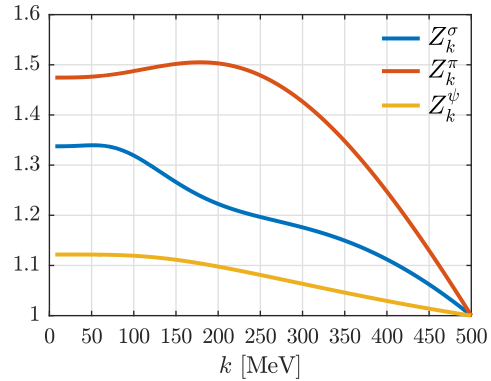}
	\caption{Scale evolution of the wave-function renormalization
	factors $Z_{k}^{\,\sigma}$, $Z_{k}^{\,\pi}$, and $Z_{k}^{\,\psi}$; 
	$k_{\mathrm{IR}} = 7$ MeV.}
	\label{fig:wave}
\end{figure}

To get an impression about the correction from unrenormalized to 
renormalized quantities, e.g.\ $m_{\sigma,k} \rightarrow M_{\sigma,k}$, 
we show the scale dependence of $Z_{k}^{\,\sigma}$, $Z_{k}^{\,\pi}$, 
and $Z_{k}^{\,\psi}$ in Fig.\ \ref{fig:wave}.
Apparently, the overall change of $Z_{k}^{\,\sigma}$ and $Z_{k}^{\,\pi}$ 
is larger compared to the one of $Z_{k}^{\,\psi}$. The former two 
already split at small $\sigma_{0,k}$ according to the evaluation of 
their flow equations at the IR minimum. The IR values of $Z_{k}^{\,\sigma}$
and $Z_{k}^{\,\pi}$ are 1.34 and 1.47, respectively, while that of $Z_{k}^{\,\psi}$ 
is 1.12. This means that the meson masses are renormalized by factors of $1/\sqrt{1.34}\simeq 0.75$
and $1/\sqrt{1.47}\simeq 0.68$, respectively, while the quark mass is renormalized 
by a factor of $1/1.12 \simeq 0.89$. Multiplying $\sigma_{0,k} = 76.9$ MeV with the 
factor $\sqrt{Z_{k}^{\,\pi}}$ yields the 
desired value of $f_{\pi}$, as mentioned above.

\begin{figure}[ht]
	\centering
		\includegraphics{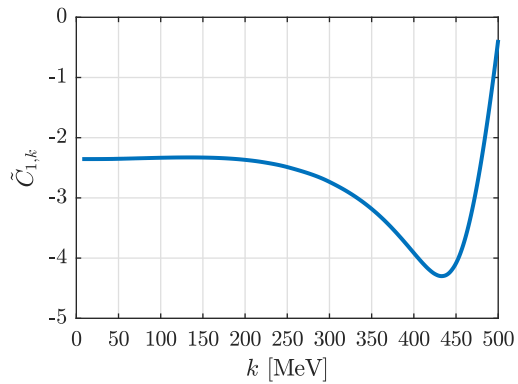}
	\caption{Scale evolution of the renormalized coupling $\tilde{C}_{1,k}$; 
	$k_{\mathrm{IR}} = 7$ MeV.}
	\label{fig:cp0}
\end{figure}

Figure\ \ref{fig:cp0} shows the scale dependence of the momentum-independent
four-pion coupling $\tilde{C}_{1,k}$. It is initialized with a value of 
$-\lambda_{\Lambda}/4=-0.3750$ at the UV cutoff (cf. Tab.\ \ref{tab:UVpars}) and 
flows to its IR value of $-2.3550$.

\begin{figure}[ht]
	\centering
		\includegraphics{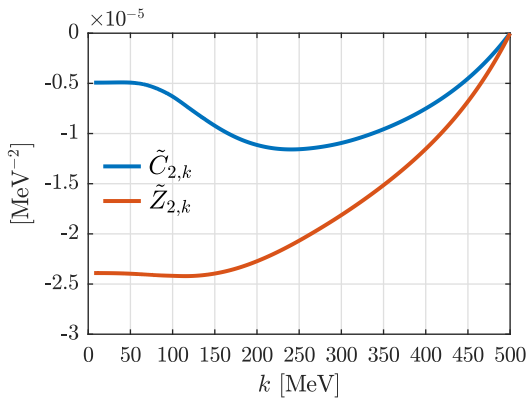}
	\caption{Scale evolution of the renormalized couplings 
	$\tilde{C}_{2,k}$ and $\tilde{Z}_{2,k}$; 
	$k_{\mathrm{IR}} = 7$ MeV.}
	\label{fig:cp2}
\end{figure}

Figure\ \ref{fig:cp2} presents the scale evolution of the derivative couplings 
$\tilde{C}_{2,k}$ and $\tilde{Z}_{2,k}$. 
Starting from an initial value of zero in the UV, these couplings 
are highly sensitive to changes of the renormalization scale $k$ 
in the region between 450 and 500 MeV. The respective curves are especially steep for 
$k\rightarrow \Lambda$. This is a clear indication that the chosen cutoff
scale is actually too low for the determination of these couplings. Unfortunately,
larger cutoffs would exceed the range of validity of the quark-meson model as
a low-energy effective theory. The fixed point-like 
behavior of $\tilde{C}_{2,k}$ and $\tilde{Z}_{2,k}$ 
below $k=250$ MeV, however, can be seen as an indication for a 
rather mild dependence of the IR values on their initial UV values. 
Nevertheless, a direct calculation from QCD along the lines
of Refs.\ \cite{Braun:2014ata, Mitter:2014wpa, Cyrol:2017ewj} 
would be preferable for their determination.

\begin{figure}[ht]
	\centering
		\includegraphics{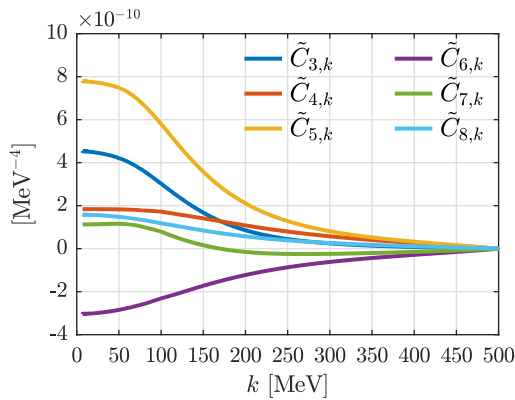}
	\caption{Scale evolution of the renormalized couplings 
	$\tilde{C}_{3,k}$ to $\tilde{C}_{8,k}$;
	$k_{\mathrm{IR}} = 7$ MeV.}
	\label{fig:cp4}
\end{figure}

In contrast, the evolution of $\tilde{C}_{3,k},\ldots , \tilde{C}_{8,k}$
appears to be much flatter in the high-energy region near the UV cutoff,
cf.\ Fig.\ \ref{fig:cp4}.
$\tilde{C}_{3,k},\ldots , \tilde{C}_{8,k}$ substantially 
grow or shrink only below 250 MeV. Thus, adding to the discussion above, 
in this case assuming a starting value of zero at the UV scale seems reasonable.
Figure\ \ref{fig:cp4} verifies our
expectation that the nontrivial running of the low-energy couplings of
order $\mathcal{O}(\partial^{4})$
with respect to the RG scale $k$ only sets in approximately below the mass 
threshold of the $\sigma$ field ($\simeq 400$ MeV).

In the IR limit, all momentum-dependent pion 
self-interactions differ from the unrenormalized couplings 
by a factor of $1/({Z_{k}^{\,\pi}})^{2} = 0.46$. The exact IR 
values of Figs.\ \ref{fig:cp0}, \ref{fig:cp2}, and \ref{fig:cp4}
are listed in the column ``trunc'' in Tab.\ \ref{tab:LECs}.
\begin{table}[h]
	\caption{\label{tab:LECs}Low-energy couplings 
	($f_{\pi} = 93$ MeV).\footnote{The values 
	in the last three columns correspond to the renormalized couplings 
	in the IR limit.}}
	\begin{ruledtabular}
		\begin{tabular}{lcccc}
		& & \multicolumn{3}{c}{\textbf{FRG}}\\
		\cline{3-5}
		\textbf{Coupling} & \textbf{Tree level} & \textbf{tree} & \textbf{trunc} & \textbf{total} \\ 
		\colrule
		$C_{1}$ & $-0.2514$ & $2.1063$ & $-2.3550$ & $-0.2487$ \\
		$C_{2}\ [1/f_{\pi}^{2}]$ & $0.4054$ & $0.3598$ & $-0.0426$ & $0.3172$\\
		$Z_{2}\ [1/f_{\pi}^{2}]$ & -- & -- & $-0.2068$ & $-0.2068$\\
		$C_{3}\ [1/f_{\pi}^{4}]\times 10^{2}$ & $1.7311$ & $1.5364$ & $3.3946$ & $4.9310$\\
		$C_{4}\ [1/f_{\pi}^{4}]\times 10^{2}$ & -- & -- & $1.3752$ & $1.3752$\\
		$C_{5}\ [1/f_{\pi}^{4}]\times 10^{2}$ & $3.4621$ & $3.0728$ & $5.8294$ & $8.9023$\\
		$C_{6}\ [1/f_{\pi}^{4}]\times 10^{2}$ & -- & -- & $-2.2697$ & $-2.2697$\\
		$C_{7}\ [1/f_{\pi}^{4}]\times 10^{2}$ & $1.7311$ & $1.5364$ & $0.8439$ & $2.3804$\\
		$C_{8}\ [1/f_{\pi}^{4}]\times 10^{2}$ & -- & -- & $1.1828$ & $1.1828$
		\end{tabular}
	\end{ruledtabular}
\end{table}
The modified tree-level contributions $\tilde{C}^{\mathrm{tree}}_{i,k}$ with
$i \in \lbrace 1, \ldots , 8 \rbrace$ and $ \tilde{Z}^{\mathrm{tree}}_{2,k}$ 
from Eqs.\ (\ref{eq:c1o4mod}) -- (\ref{eq:co4mod}) are collected in 
the column ``tree''.  Their sums $\tilde{C}^{\mathrm{total}}_{i,k}$,
$i \in \lbrace 1, \ldots , 8 \rbrace$ and $ \tilde{Z}^{\mathrm{total}}_{2,k}$ 
are listed in the last column.

For the sake of providing a complete overview, Tab.\ \ref{tab:LECs} also 
contains (in column two) the numerical tree-level 
couplings $C_{1,O(4)}$, $C_{2,O(4)}$, and $C_{3,O(4)}$ from 
Eqs.\ (\ref{eq:c1o4}), (\ref{eq:c2o4}), and (\ref{eq:c3o4}), as well as 
$C_{5,O(4)}$ and $C_{7,O(4)}$. In contrast to Ref.\ \cite{Divotgey:2016pst},
the latter two are also generated within the chosen basis set (\ref{eq:basis2}) and (\ref{eq:basis4}). 
$Z_{2}$, $C_{4}$, $C_{6}$,
and $C_{8}$ vanish at tree-level. All tree-level results are 
produced by taking the masses and the vacuum 
expectation value in the IR as input, i.e., $m_{\sigma} = 450.1$ MeV, 
$m_{\pi} = 139.3$ MeV, and $\sigma_{0} = 93.4$ MeV.

The tree-level estimate for $C_{1} \simeq -0.2514$ is a good
approximation. The loop-corrected coupling $\tilde{C}_{1,k}^{\mathrm{total}}$ 
is only by 1.1 percent larger. The large difference between the entries in the 
``Tree level'' and ``tree'' columns for $C_{1}$ arises from the fact that
the ``Tree level'' value contains both the contribution from integrating out the $\sigma$ field
as well as that from the four-pion interaction
in the potential. On the other hand, in the case of the FRG calculation the ``tree'' value only 
contains the contribution from integrating out the $\tilde{\sigma}$ field,
while the ``trunc'' value arises from the potential $\tilde{U}_k$.

For the couplings of order 
$\mathcal{O}(\partial^{2})$ and $\mathcal{O}(\partial^{4})$, 
we notice that the tree-level contributions slightly decrease 
on account of the correction from the wave-function renormalization 
factors (second column compared to the third one). For
$C_{2}$ we find a negative value in the column ``trunc'', but all other
values in the same row are positive.
The total FRG results differ from the tree-level estimates
by a factor of 0.78, 2.85, 2.57, and 1.38 for $C_{2}$, $C_{3}$, $C_{5}$,
and $C_{7}$, respectively. Hence, the corrections
due to loop contributions are smallest for the momentum-independent $C_{1}$ coupling.

\begin{figure*}[t]
	\centering
		\includegraphics{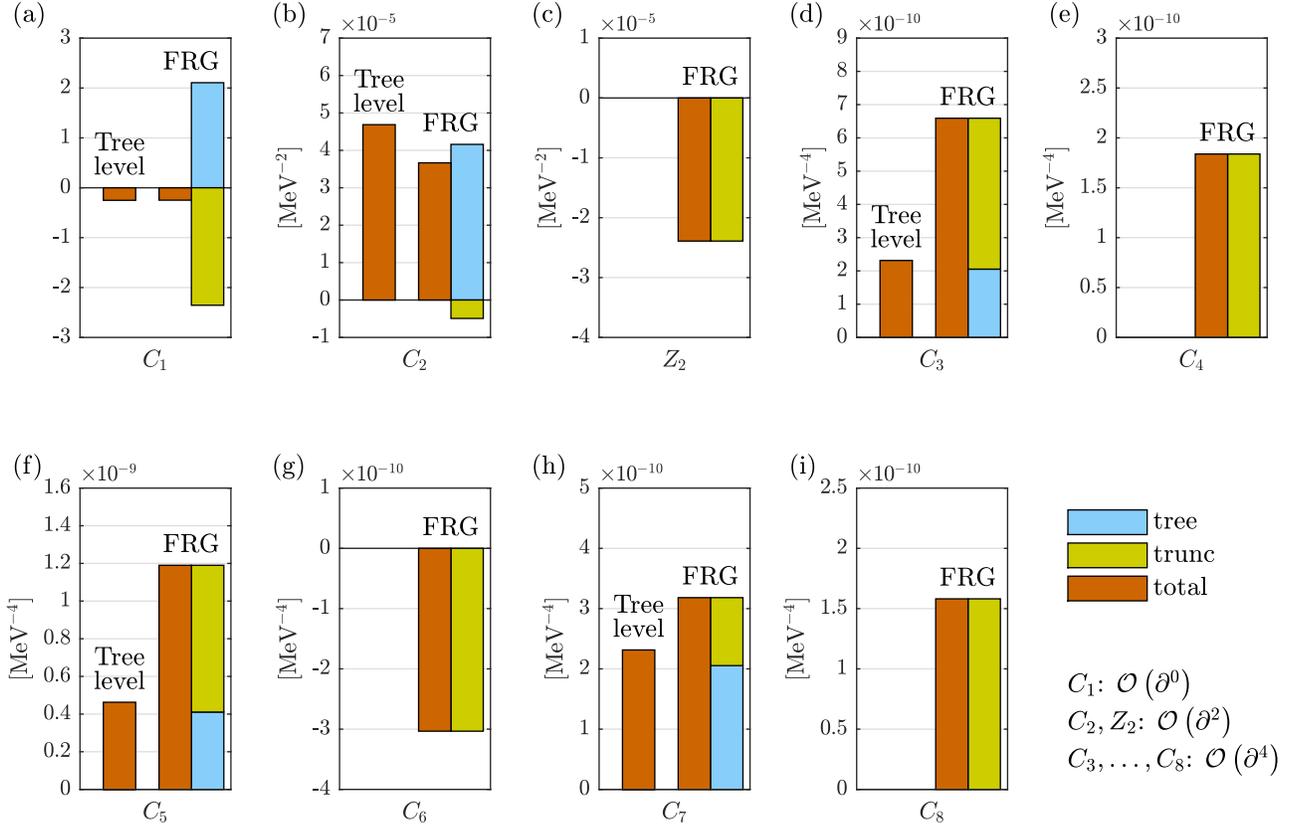}
	\caption{Comparison of the low-energy couplings $C_{1}, \ldots , C_{8}$, and $Z_{2}$ 
	obtained from the tree-level estimate and the 
	FRG calculation. The contributions to the different 
	couplings are given in separate subfigures (a) -- (i). The order
	in the derivative expansion is summarized in the lower right corner. 
	The bar heights correspond to the numerical
	values quoted in Tab.\ \ref{tab:LECs}. The $k$-dependent 
	quantities are evaluated at $k_{\mathrm{IR}} = 7$ MeV.}
	\label{fig:cbar}
\end{figure*}

Figure \ref{fig:cbar} visualizes Tab.\ \ref{tab:LECs}
in terms of a bar plot.
In subfigures 
(a), (b), (d), and (h), both results are of the same order of magnitude 
($\sim 10^{-1}$ for $C_{1}$, $\sim 10^{-5}\ \mathrm{MeV}^{-2}$ for $C_{2}$, and 
$\sim 10^{-10}\ \mathrm{MeV}^{-4}$ for $C_{3}$, $C_{7}$), but this does not hold
for $C_{5}$ in subfigure (f).

\begin{figure}[ht]
	\centering
		\includegraphics{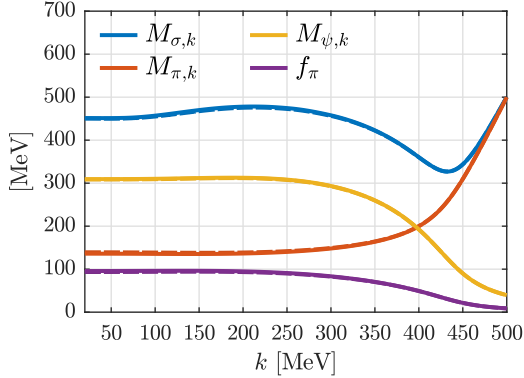}
	\caption{Scale evolution of the renormalized meson and quark masses 
	within the LPA' truncation. The dashed lines show the related results from
	Fig.\ \ref{fig:masses}; $k_{\mathrm{IR}} = 20$ MeV.}
	\label{fig:masses_LPAp}
\end{figure}

As the last point in this study, the amount of feedback 
of the scale evolution of $C_{2,k}$, $Z_{2,k}$, and $C_{3,k}, \ldots , C_{8,k}$ 
onto the masses and the wave-function renormalization is 
assessed by comparing the presented results to the LPA' flow.
The latter is initialized with the same UV parameters as in Tab.\ \ref{tab:UVpars}. 
Here, the higher-derivative couplings remain zero for all $k$. 
Within the LPA' truncation, the qualitative behavior of 
the observables in Fig.\ \ref{fig:masses} is reproduced, 
cf. Fig.\ \ref{fig:masses_LPAp}.
The IR values (compared to the truncation beyond LPA' including the higher 
couplings) are found to be: $M_{\sigma,k} = 451.5$ (450.1) MeV, $M_{\pi,k} = 136.1$ 
(139.3) MeV, $M_{\psi,k} = 309.4$ (308.5) MeV, and 
$\sigma_{0,k}\sqrt{Z_{k}^{\,\pi}} = 95.7$ (93.4) MeV.

\begin{figure}[ht]
	\centering
		\includegraphics{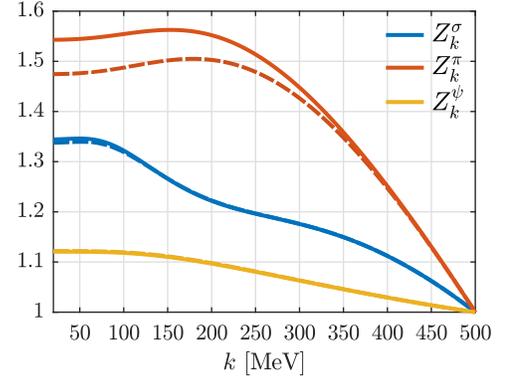}
	\caption{Scale evolution of the wave-function renormalization
	factors $Z_{k}^{\,\sigma}$, $Z_{k}^{\,\pi}$, and $Z_{k}^{\,\psi}$
	within the LPA' truncation. The dashed lines show the related results from
	Fig.\ \ref{fig:wave}; $k_{\mathrm{IR}} = 20$ MeV.}
	\label{fig:wave_LPAp}
\end{figure}

At the same time, Fig.\ \ref{fig:wave_LPAp} reveals a significant change in 
the evolution of $Z_{k}^{\,\pi}$ within the LPA', whereas $Z_{k}^{\,\sigma}$ and 
$Z_{k}^{\,\psi}$ are only slightly affected: $Z_{k}^{\,\pi} = 1.54$ (1.47), 
$Z_{k}^{\,\sigma} = 1.34$ (1.34), and $Z_{k}^{\,\psi} = 1.12$ (1.12) at 
$k_{\mathrm{IR}} = 20$ MeV.
The reason for this comparatively large interference of the flow of $Z_{k}^{\,\pi}$ 
with that of $C_{2,k}$, $Z_{2,k}$, and $C_{3,k}, \ldots , C_{8,k}$ is immediately deduced 
from Eqs.\ (\ref{eq:zs}) -- (\ref{eq:zf}) in Appendix\ \ref{sec:floweqns}: 
The flow of $Z_{k}^{\,\pi}$ solely involves a four-point pion vertex, which we do not 
find on the right-hand sides of Eqs.\ (\ref{eq:zs}) and (\ref{eq:zf}). 
This leads to a direct coupling of the 
pion wave-function renormalization to the momentum-dependent pion self-interactions. 
The other factors $Z_{k}^{\,\sigma}$ and $Z_{k}^{\,\psi}$ are only indirectly coupled.

\section{Summary and Outlook}
\label{sec:summary}

In this work, we have computed the low-energy limit of 
the $O(4)$-symmetric quark-meson model as a 
limit of the eLSM. In particular, we calculated quantum 
corrections to the momentum-dependent four-pion interactions. 
As discussed in Secs.\ \ref{sec:FRG} and \ref{sec:results}
neglecting $\sigma$-dependent derivative interactions
within the fully $O(4)$-symmetric model corresponds to an 
approximation validated by the results shown in 
Fig.\ \ref{fig:cp4}.

We applied the FRG approach and compared the 
obtained results to a tree-level estimate. 
In the latter approximation, contributions to these pion 
self-interactions partly arise from a mapping of the low-energy effective 
action of the eLSM onto the one of ChPT, see Ref.\ \cite{Divotgey:2016pst}.
We extended these contributions to a full set of basis structures for the orders
$\mathcal{O}(\partial^{2})$ and $\mathcal{O}(\partial^{4})$.

The quantum corrections are due to the exchange of the quark as well as 
the meson degrees of freedom of the quark-meson model as the low-energy 
effective theory of QCD. They have been calculated in the 
nonperturbative FRG approach, where the higher-derivative couplings have 
been treated as additional terms in a truncation beyond the well-known LPA' 
approximation of the quark-meson model.

We found that the overall order of magnitude of $C_{1}$, $C_{2}$, $C_{3}$, and
$C_{7}$ is already set by their tree-level estimates, cf. Fig.\ \ref{fig:cbar}.
The loop-corrected values of $C_{1}$, $C_{2}$, and $C_{7}$ only differ by 
factors of 0.99, 0.78, and 1.38 from the tree-level estimates. 
However, the remaining couplings $Z_{2}$, $C_{4}$, $C_{5}$, $C_{6}$, 
and $C_{8}$ seem to be dominated by the loop corrections that 
we obtained from the FRG.

We also found a qualitative difference in the origin of 
the quantum contributions to $C_{2}$ and $Z_{2}$ on the one hand
and $C_{3}, \ldots , C_{8}$ on the other hand. Whereas the former 
already receive large contributions at the initial cutoff, which are 
dominated by quark fluctuations, the quantum corrections to the latter are
created at low RG scales, i.e., the regime that is governed by pion fluctuations. 
These findings encourage a study within full QCD along the 
lines of Refs.\ \cite{Mitter:2014wpa, Braun:2014ata, Cyrol:2017ewj} 
to determine the low-energy couplings, in particular $C_{2}$ 
and $Z_{2}$, from first-principles QCD.

On a more technical level it was demonstrated that, although the 
wave-function renormalization of the pion field significantly 
shrinks when we extend the LPA' by the higher couplings 
$C_{2,k}$, $Z_{2,k}$, and $C_{3,k}, \ldots , C_{8,k}$, the 
$k$-dependent meson and quark masses are rather robust 
against such an extension of the truncation.

It needs to be clarified how the behavior of $C_{2}$ and $Z_{2}$,
as well as $C_{3}, \ldots , C_{8}$ would change if 
we consider the fully $O(4)$-symmetric scenario, i.e., 
including momentum-dependent $\sigma$ self-interactions and 
$\sigma\pi$ couplings. Such a study is, however, beyond the scope of 
the present work. Moreover, in principle one has to check whether 
these results are reproducible under variations or extensions of 
the regulators (\ref{eq:reg1}) -- (\ref{eq:reg3}), as studied
in Ref.\ \cite{Pawlowski:2015mlf}, and the truncation 
(\ref{eq:specansatz}). Especially, the last point would be 
interesting in the context of the running of the low-energy 
couplings. In addition, it would be important to allow for 
a scale-dependence of the Yukawa coupling, because it is closely 
related to the quark mass and the pion decay constant. All of 
these points will be addressed in a future investigation that 
incorporates also effects from full QCD \cite{Cyrol:2017ewj}, 
where the dynamical hadronization technique has been applied 
to describe quark-antiquark bound states as mesonic degrees of 
freedom.

Finally, as repeatedly mentioned, the $O(4)$ quark-meson model is only 
the specific limit of the eLSM where (axial\hbox{-})\-vector 
mesons are neglected. Nevertheless, the ultimate goal of subsequent 
studies remains a comparison of the LECs of the full eLSM to 
those of ChPT. It might be therefore necessary to extend the ansatz 
(\ref{eq:genansatz}) to the full effective action of the eLSM and,
furthermore, to translate the presented
projection of the momentum-dependence of the four-pion
vertex to the basis that naturally arises within the chiral expansion
of ChPT.

Another important check would be to analyze the pion-mass dependence 
of the low-energy couplings produced within the FRG, thus extending 
the work of Ref.\ \cite{Jendges:2006yk} to
higher-derivative couplings. This would 
constitute a nontrivial confirmation that the correct IR behavior 
is respected by the truncation introduced in this work, 
which corresponds to a partial resummation of loop diagrams.

\begin{acknowledgments}
We are very grateful to S.\ Rechenberger for helpful advice and 
many enlightening comments on this work. We also would like 
to thank M.\ Birse, J.\ Braun, D.D.\ Dietrich, A.\ Koenigstein, 
J.M.\ Pawlowski, R.D.\ Pisarski, F.\ Rennecke, B.-J.\ Schaefer, 
N.\ Strodthoff, and L.\ von Smekal
for valuable discussions. J.E.\ acknowledges funding by the German 
National Academic Foundation. M.M.\ acknowledges funding by the FWF 
grant J3507-N27, the DFG grant MI 2240/1-1 and the U.S.\ Department 
of Energy under contract de-sc0012704. D.H.R.\ is partially supported by the High-end Foreign 
Experts project GDW20167100136 of the State
Administration of Foreign Experts Affairs of China. 
\end{acknowledgments}

\appendix

\section{PCAC}
\label{sec:PCAC}

In this Appendix we present further details on the PCAC 
relation used in Eq.\ (\ref{eq:PCAC}). First, we consider 
the behavior of the quark fields under infinitesimal $SU(2)_{A}$ 
transformations,
\begin{IEEEeqnarray}{rCl}
	\psi & \rightarrow & U\psi = \psi - i\alpha_{A,a}
	\gamma_{5}t_{a}\psi
	 = \psi + \delta\psi, \\
	\bar{\psi} & \rightarrow & \bar{\psi} U = 
	\bar{\psi} - i\bar{\psi} \alpha_{A,a}\gamma_{5}t_{a}
	 = \bar{\psi} + \delta\bar{\psi},
\end{IEEEeqnarray}
with $U \in SU(2)_{A}$,
\begin{equation}
	U = \exp\left(-i \alpha_{A,a}\gamma_{5}t_{a}\right),
	\quad a = 1,2,3 .
\end{equation}
From the Yukawa interaction in Eq.\ (\ref{eq:specansatz}), 
we deduce the transformation behavior of $\Phi_{5}$,
\begin{equation}
	\Phi_{5} \rightarrow \Phi_{5} + i\alpha_{A,a}\gamma_{5}
	t_{a}\left(\sigma + \sigma_{0}\right)
	- \alpha_{A,a} \pi_{a} t_{0}.
\end{equation}
As a consequence, we find
\begin{equation}
	\delta\sigma = - \alpha_{A,a}\pi_{a},\qquad \delta\pi_{a} 
	= \alpha_{A,a}\left(\sigma + \sigma_{0}\right).
\end{equation}
In terms of renormalized fields, these transformations read
\begin{IEEEeqnarray}{rCl}
	\delta\tilde{\psi} & = & -i \alpha_{A,a}\gamma_{5}t_{a} 
	\tilde{\psi}, \\
	\delta\tilde{\bar{\psi}} & = & -i \tilde{\bar{\psi}} 
	\alpha_{A,a}\gamma_{5}t_{a}, \\
	\delta\tilde{\sigma} & = & -\sqrt{Z_{k}^{\,\sigma}} 
	\alpha_{A,a} \frac{\tilde{\pi}_{a}}{\sqrt{Z_{k}^{\,\pi}}}, \\
	\delta\tilde{\pi}_{a} & = & \sqrt{\frac{Z_{k}^{\,\pi}}{Z_{k}^{\,\sigma}}} \alpha_{A,a} 
	\left(\tilde{\sigma} + \tilde{\sigma}_{0} \right).
\end{IEEEeqnarray}
The axial-vector current in Minkowski space is then given by
($g_{\mu\nu}$ denotes the metric tensor)
\begin{IEEEeqnarray}{rCl}
	\mathcal{J}_{\mu,a}^{A} & = & \tilde{\bar{\psi}} 
	\gamma_{\mu} \gamma_{5} t_{a} \tilde{\psi}
	 - \left(\partial_{\nu} \tilde{\sigma}\right) 
	 \sqrt{\frac{Z_{k}^{\,\sigma}}{Z_{k}^{\,\pi}}} 
	\bigg(\tilde{\pi}_{a} g_{\mu}^{\ \nu} \nonumber\\ 
	& & \qquad - \frac{Z_{k}^{\,\pi}}{Z_{k}^{\,\sigma}}
	\bigg\lbrace \tilde{C}_{5,k} \left[\left(\partial_{\rho}
	\tilde{\vec{\pi}}\right)
	\cdot \partial^{\rho}\tilde{\vec{\pi}}\right]
	\tilde{\pi}_{a} g_{\mu}^{\ \nu} \nonumber\\
	& & \qquad\qquad\quad +\, 2\, \tilde{C}_{6,k}\, \tilde{\vec{\pi}}^{2}
	\partial_{\mu}\partial^{\nu}\tilde{\pi}_{a} 
	\nonumber\\
	& & \qquad\qquad\quad +\, 2\, \tilde{C}_{7,k} \left(\tilde{\vec{\pi}}
	\cdot \partial_{\rho}\partial^{\rho}\tilde{\vec{\pi}}\right)
	\tilde{\pi}_{a} g_{\mu}^{\ \nu} \nonumber\\
	& & \qquad\qquad\quad +\, 2\, \tilde{C}_{8,k}\, \tilde{\vec{\pi}}^{2}
	\partial_{\rho}\partial^{\rho}\tilde{\pi}_{a} g_{\mu}^{\ \nu}
	\bigg\rbrace \bigg) \nonumber\\
	& & + \sqrt{\frac{Z_{k}^{\,\pi}}{Z_{k}^{\,\sigma}}}\left(\tilde{\sigma}
	+ \tilde{\sigma}_{0}\right) \nonumber\\
	& & \quad \times \bigg\lbrace\partial_{\mu}\tilde{\pi}_{a} 
	+ 2\, \tilde{C}_{2,k}\left(\tilde{\vec{\pi}} \cdot
	\partial_{\mu} \tilde{\vec{\pi}}\right) 
	\tilde{\pi}_{a} + 2\, \tilde{Z}_{2,k}\, \tilde{\vec{\pi}}^{2}
	\partial_{\mu}\tilde{\pi}_{a} \nonumber\\
	& & \quad \quad \ + \left(4\, \tilde{C}_{3,k}
	- \tilde{C}_{5,k}\right) \left(\partial_{\nu}
	\tilde{\vec{\pi}}\right)^{2}
	\partial_{\mu}\tilde{\pi}_{a} \nonumber\\
	& & \quad \quad \ +\, 4\, \tilde{C}_{4,k}
	\left[\left(\partial_{\mu}\tilde{\vec{\pi}}\right)
	\cdot \partial_{\nu}\tilde{\vec{\pi}}\right]
	\partial^{\nu}\tilde{\pi}_{a} \nonumber\\
	& & \quad \quad \ +\, 2\left(\tilde{C}_{5,k}
	- \tilde{C}_{7,k}\right)
	\left(\tilde{\vec{\pi}}
	\cdot \partial_{\nu}\partial^{\nu}\tilde{\vec{\pi}}\right)
	\partial_{\mu}\tilde{\pi}_{a} \nonumber\\
	& & \quad \quad \ -\, 2\, \tilde{C}_{5,k}
	\left[\left(\partial_{\mu}\partial_{\nu}\tilde{\vec{\pi}}\right)
	\cdot \partial^{\nu}\tilde{\vec{\pi}}\right]
	\tilde{\pi}_{a} \nonumber\\
	& & \quad \quad \ -\, 4\, \tilde{C}_{6,k}
	\left(\tilde{\vec{\pi}}
	\cdot \partial_{\nu}\tilde{\vec{\pi}}\right)
	\partial_{\mu}\partial^{\nu}\tilde{\pi}_{a} \nonumber\\
	& & \quad \quad \ -\, 2\left(\tilde{C}_{6,k}
	+ \tilde{C}_{8,k}\right) \tilde{\vec{\pi}}^{2}
	\partial_{\mu}\partial_{\nu}\partial^{\nu}\tilde{\pi}_{a} \nonumber\\
	& & \quad \quad \ -\, 2\, \tilde{C}_{7,k}
	\left[\left(\partial_{\mu}\tilde{\vec{\pi}}\right)
	\cdot \partial_{\nu}\partial^{\nu}\tilde{\vec{\pi}}\right]
	\tilde{\pi}_{a} \nonumber\\
	& & \quad \quad \ -\, 2\, \tilde{C}_{7,k}
	\left(\tilde{\vec{\pi}}\cdot \partial_{\mu}
	\partial_{\nu}\partial^{\nu}\tilde{\vec{\pi}}\right)
	\tilde{\pi}_{a} \nonumber\\
	& & \quad \quad \ -\, 4\, \tilde{C}_{8,k}
	\left(\tilde{\vec{\pi}}\cdot \partial_{\mu}
	\tilde{\vec{\pi}}\right)
	\partial_{\nu}\partial^{\nu}\tilde{\pi}_{a}	
	\bigg\rbrace .
\end{IEEEeqnarray}

\section{Flow equations}
\label{sec:floweqns}

We derive the flow equations for all scale-dependent parts on the 
right-hand side of Eq.\ (\ref{eq:specansatz}) from Eq.\ (\ref{eq:Wetterich}) 
and functional derivatives thereof. To this end, we partly transform 
the effective average action (\ref{eq:specansatz}) into momentum space,
\begin{IEEEeqnarray}{rCl}
	\Gamma_{k} & = & \mathcal{V}\sum_{q}\bigg[\frac{Z_{k}^{\,\sigma}}{2}
	q^{2} \sigma(-q)\sigma(q) + \frac{Z_{k}^{\,\pi}}{2} q^{2}
	\vec{\pi}(-q) \cdot \vec{\pi}(q) \nonumber\\
	& & \quad \qquad +\, \bar{\psi}(q)\left(-i 
	Z_{k}^{\,\psi}\gamma_{\mu}q_{\mu}\right)\psi(q)\bigg] \nonumber\\ 
	& & + \int_{x}\left[U_{k}(\rho)-h_{\mathrm{ESB}}
	\sigma\right] \nonumber\\
	& & - \mathcal{V}\sum_{q_{1},q_{2},q_{3}} 
	\big\lbrace C_{2,k}\, q_{1,\mu} q_{3,\mu} 
	+ Z_{2,k}\, q_{2,\mu} q_{3,\mu} \nonumber\\
	& & \qquad \qquad \quad -\, C_{3,k}\, (q_{1}+q_{2}
	+q_{3})_{\mu} q_{1,\mu}\, q_{2,\nu} q_{3,\nu} \nonumber\\
	& & \qquad \qquad \quad -\, C_{4,k}\, (q_{1}+q_{2}
	+q_{3})_{\mu} q_{2,\mu}\, q_{1,\nu} q_{3,\nu} \nonumber\\
	& & \qquad \qquad \quad -\, C_{5,k}\, (q_{1}+q_{2}
	+q_{3})_{\mu} q_{1,\mu}\, q_{3}^{2} \nonumber\\
	& & \qquad \qquad \quad +\, C_{6,k}\, \left[(q_{1}+q_{2}
	+q_{3})_{\mu} q_{1,\mu}\right]^{2} \nonumber\\
	& & \qquad \qquad \quad +\, C_{7,k}\, q_{1}^{2}
	\, q_{3}^{2} \nonumber\\
	& & \qquad \qquad \quad +\, C_{8,k}\, (q_{1}+q_{2}
	+q_{3})^{2} q_{1}^{2}
	\big\rbrace
	\nonumber\\
	& & \qquad \quad \times \
	\pi_{i}(-q_{1}-q_{2}-q_{3}) 
	\pi_{i}(q_{1})
	\pi_{j}(q_{2}) \pi_{j}(q_{3}) \nonumber\\
	& & + \mathcal{V}\sum_{q_{1}, q_{2}} y\bar{\psi}(q_{1})
	\Phi_{5}(q_{1}-q_{2})\psi(q_{2}).
	\label{eq:momentumspace}
\end{IEEEeqnarray}
Isospin indices $i$ and $j$ appearing twice are summed over. They replace 
the inner products in lines three to seven of Eq.\ (\ref{eq:specansatz}).

The regulators $R_{k}$ as a function of the internal momentum $q$, typically
$R_{k}(q^{2})$ for bosonic and $R_{k}(q)$ for fermionic fields,
have to fulfill the requirements
\begin{IEEEeqnarray}{rCl}
	R_{k} \rightarrow 0 & \ \text{for}\ & k \rightarrow 0, \\
	R_{k} \rightarrow \infty & \ \text{for}\ & k \rightarrow 
	\Lambda \rightarrow \infty , \\
	R_{k} > 0 & \ \text{for}\ & q \rightarrow 0, \\
	R_{k} \rightarrow 0 & \ \text{for}\ & q \rightarrow \infty .
\end{IEEEeqnarray}
In this way, the high-energy modes ($q^{2} \gg k^{2}$) are successively 
integrated out by lowering the scale $k$. For the soft modes ($q^{2} 
\ll k^{2}$), in contrast, the regulators generate an additional mass 
contribution. This leads to an exclusion of these modes from the
integration process.

For our analysis we take exponential-type regulators
\ \cite{Wetterich:1989xg, Wetterich:1992yh, Litim:2001dt, Nandori:2012tc, 
Pawlowski:2015mlf}, namely,
\begin{IEEEeqnarray}{rCl}
	R_{k}^{\sigma}\left(q^{2}\right) & = & Z_{k}^{\,\sigma} 
	q^{2} r_{\mathrm{B}}
	\left(\frac{q^{2}}{k^{2}}\right), \label{eq:reg1}\\
	R_{k}^{\pi}\left(q^{2}\right) & = & Z_{k}^{\,\pi} q^{2} 
	r_{\mathrm{B}}
	\left(\frac{q^{2}}{k^{2}}\right), \label{eq:reg2}\\
	R_{k}^{\psi}\left(q\right) & = & -iZ_{k}^{\,\psi} 
	\gamma_{\mu} q_{\mu} r_{\mathrm{F}}
	\left(\frac{q^{2}}{k^{2}}\right), \label{eq:reg3}
\end{IEEEeqnarray}
\begin{widetext}
where the bosonic and fermionic shape functions $r_{\mathrm{B}}$ 
and $r_{\mathrm{F}}$ are
\begin{IEEEeqnarray}{rCl}
	r_{\mathrm{B}}(x) & = & \frac{1}{\exp\left(x^{2}\right) - 1}, \\
	r_{\mathrm{F}}(x) & = & \sqrt{1 + r_{\mathrm{B}}(x)} - 1.
\end{IEEEeqnarray}

Acting now with the derivatives on $\Gamma_{k}$ and using 
the diagrammatic representations from Eq.\ (\ref{eq:rep}), 
the flow equations for the effective potential $U_{k}$, 
the wave-function renormalizations $Z_{k}^{\,\sigma}$, $Z_{k}^{\,\pi}$, 
and $Z_{k}^{\,\psi}$, as well as the higher couplings $C_{2,k}$,
$Z_{2,k}$, and $C_{3,k}$, $\ldots$ , $C_{8,k}$ take the form
\begin{IEEEeqnarray}{rCl}
	\partial_{k}U_{k} & = & \mathcal{V}^{-1} \partial_{k}\Gamma_{k}
	 = \mathcal{V}^{-1} \biggg( 
	\frac{1}{2} \! \! \vcenter{\hbox{
	\begin{pspicture}[showgrid=false](1.5,2.4)
		\psarc[linewidth=0.02,linestyle=dashed,dash=2pt 1pt,linecolor=blue](0.75,1.2){0.6}{115}{65}
		\pscircle[linewidth=0.03,fillstyle=solid,fillcolor=NavyBlue](0.75,1.8){0.25}
		\psline[linewidth=0.03](0.75,1.8)(0.92,1.97)
		\psline[linewidth=0.03](0.75,1.8)(0.58,1.97)
		\psline[linewidth=0.03](0.75,1.8)(0.58,1.63)
		\psline[linewidth=0.03](0.75,1.8)(0.92,1.63)
		\rput[b]{*0}(0.75,0.35){$\sigma$}
	\end{pspicture}
	}} \! \! \!
	+ \frac{1}{2} \! \!
	\vcenter{\hbox{
	\begin{pspicture}(1.5,2.4)
		\psarc[linewidth=0.02,linestyle=dashed,dash=2pt 1pt,linecolor=red](0.75,1.2){0.6}{115}{65}
		\pscircle[linewidth=0.03,fillstyle=solid,fillcolor=RedOrange](0.75,1.8){0.25}
		\psline[linewidth=0.03](0.75,1.8)(0.92,1.97)
		\psline[linewidth=0.03](0.75,1.8)(0.58,1.97)
		\psline[linewidth=0.03](0.75,1.8)(0.58,1.63)
		\psline[linewidth=0.03](0.75,1.8)(0.92,1.63)
		\rput[b]{*0}(0.75,0.35){$\pi$}
	\end{pspicture}
	}} \! \! \!
	- \! \! \!
	\vcenter{\hbox{
	\begin{pspicture}(1.5,2.4)
		\psarc[linewidth=0.02,arrowsize=2pt 3,arrowinset=0]{->}(0.75,1.2){0.6}{115}{189}
		\psarc[linewidth=0.02](0.75,1.2){0.6}{180}{65}
		\pscircle[linewidth=0.03,fillstyle=solid,fillcolor=gray](0.75,1.8){0.25}
		\psline[linewidth=0.03](0.75,1.8)(0.92,1.97)
		\psline[linewidth=0.03](0.75,1.8)(0.58,1.97)
		\psline[linewidth=0.03](0.75,1.8)(0.58,1.63)
		\psline[linewidth=0.03](0.75,1.8)(0.92,1.63)
		\rput[b]{*0}(0.75,0.2){$\psi$}
	\end{pspicture}
	}} \! \! \biggg), \\
	\partial_{k}Z_{k}^{\,\sigma} & = & \mathcal{V}^{-1} 
	\left.\frac{\mathrm{d}}{\mathrm{d}p^2}\right|_{p^{2}=0}
	\frac{\delta^{2}\partial_{k}\Gamma_{k}}{\delta\sigma(-p)\delta\sigma(p)} \nonumber\\
	& = & \mathcal{V}^{-1} \left.\frac{\mathrm{d}}{\mathrm{d}p^2}\right|_{p^{2}=0}\left(
	\frac{1}{2} \! \! \vcenter{\hbox{
	\begin{pspicture}[showgrid=false](3.0,2.0)
		\psarc[linewidth=0.02,linestyle=dashed,dash=2pt 1pt,linecolor=blue](1.5,1.0){0.6}{20}{65}
		\psarc[linewidth=0.02,linestyle=dashed,dash=2pt 1pt,linecolor=blue](1.5,1.0){0.6}{115}{160}
		\psarc[linewidth=0.02,linestyle=dashed,dash=2pt 1pt,linecolor=blue](1.5,1.0){0.6}{200}{340}
		\psline[linewidth=0.02,linestyle=dashed,dash=2pt 1pt,linecolor=blue](0.1,1.0)(0.7,1.0)
		\psline[linewidth=0.02,linestyle=dashed,dash=2pt 1pt,linecolor=blue](2.3,1.0)(2.9,1.0)
		\pscircle[linewidth=0.03,fillstyle=solid,fillcolor=NavyBlue](1.5,1.6){0.25}
		\psline[linewidth=0.03](1.5,1.6)(1.67,1.77)
		\psline[linewidth=0.03](1.5,1.6)(1.33,1.77)
		\psline[linewidth=0.03](1.5,1.6)(1.33,1.43)
		\psline[linewidth=0.03](1.5,1.6)(1.67,1.43)
		\pscircle[linewidth=0.03,fillstyle=solid,fillcolor=lightgray](0.9,1.0){0.20}
		\pscircle[linewidth=0.03,fillstyle=solid,fillcolor=lightgray](2.1,1.0){0.20}
		\rput[b]{*0}(2.1,1.4){$\sigma$}
		\rput[b]{*0}(0.9,1.4){$\sigma$}
		\rput[b]{*0}(0.4,0.7){$\sigma$}
		\rput[b]{*0}(2.6,0.7){$\sigma$}
		\rput[b]{*0}(1.5,0.15){$\sigma$}
		\rput[b]{*0}(0.9,0.92){\scriptsize{$3$}}
		\rput[b]{*0}(2.1,0.92){\scriptsize{$3$}}
	\end{pspicture}
	}} \! \! \! 
	 + \frac{1}{2} \! \! \vcenter{\hbox{
	\begin{pspicture}(3.0,2.0)
		\psarc[linewidth=0.02,linestyle=dashed,dash=2pt 1pt,linecolor=red](1.5,1.0){0.6}{20}{65}
		\psarc[linewidth=0.02,linestyle=dashed,dash=2pt 1pt,linecolor=red](1.5,1.0){0.6}{115}{160}
		\psarc[linewidth=0.02,linestyle=dashed,dash=2pt 1pt,linecolor=red](1.5,1.0){0.6}{200}{340}
		\psline[linewidth=0.02,linestyle=dashed,dash=2pt 1pt,linecolor=blue](0.1,1.0)(0.7,1.0)
		\psline[linewidth=0.02,linestyle=dashed,dash=2pt 1pt,linecolor=blue](2.3,1.0)(2.9,1.0)
		\pscircle[linewidth=0.03,fillstyle=solid,fillcolor=RedOrange](1.5,1.6){0.25}
		\psline[linewidth=0.03](1.5,1.6)(1.67,1.77)
		\psline[linewidth=0.03](1.5,1.6)(1.33,1.77)
		\psline[linewidth=0.03](1.5,1.6)(1.33,1.43)
		\psline[linewidth=0.03](1.5,1.6)(1.67,1.43)
		\pscircle[linewidth=0.03,fillstyle=solid,fillcolor=lightgray](0.9,1.0){0.20}
		\pscircle[linewidth=0.03,fillstyle=solid,fillcolor=lightgray](2.1,1.0){0.20}
		\rput[b]{*0}(2.1,1.4){$\pi$}
		\rput[b]{*0}(0.9,1.4){$\pi$}
		\rput[b]{*0}(0.4,0.7){$\sigma$}
		\rput[b]{*0}(2.6,0.7){$\sigma$}
		\rput[b]{*0}(1.5,0.15){$\pi$}
		\rput[b]{*0}(0.9,0.92){\scriptsize{$3$}}
		\rput[b]{*0}(2.1,0.92){\scriptsize{$3$}}
	\end{pspicture}
	}} \! \! \!  
	- \! \! \! \vcenter{\hbox{
	\begin{pspicture}(3.0,2.0)
		\psarc[linewidth=0.02,arrowsize=2pt 3,arrowinset=0]{->}(1.5,1.0){0.6}{20}{55}
		\psarc[linewidth=0.02](1.5,1.0){0.6}{46}{65}
		\psarc[linewidth=0.02,arrowsize=2pt 3,arrowinset=0]{->}(1.5,1.0){0.6}{115}{150}
		\psarc[linewidth=0.02](1.5,1.0){0.6}{141}{160}
		\psarc[linewidth=0.02,arrowsize=2pt 3,arrowinset=0]{->}(1.5,1.0){0.6}{200}{279}
		\psarc[linewidth=0.02](1.5,1.0){0.6}{270}{340}
		\psline[linewidth=0.02,linestyle=dashed,dash=2pt 1pt,linecolor=blue](0.1,1.0)(0.7,1.0)
		\psline[linewidth=0.02,linestyle=dashed,dash=2pt 1pt,linecolor=blue](2.3,1.0)(2.9,1.0)
		\pscircle[linewidth=0.03,fillstyle=solid,fillcolor=gray](1.5,1.6){0.25}
		\psline[linewidth=0.03](1.5,1.6)(1.67,1.77)
		\psline[linewidth=0.03](1.5,1.6)(1.33,1.77)
		\psline[linewidth=0.03](1.5,1.6)(1.33,1.43)
		\psline[linewidth=0.03](1.5,1.6)(1.67,1.43)
		\pscircle[linewidth=0.03,fillstyle=solid,fillcolor=lightgray](0.9,1.0){0.20}
		\pscircle[linewidth=0.03,fillstyle=solid,fillcolor=lightgray](2.1,1.0){0.20}
		\rput[b]{*0}(2.1,1.4){$\psi$}
		\rput[b]{*0}(0.9,1.4){$\psi$}
		\rput[b]{*0}(0.4,0.7){$\sigma$}
		\rput[b]{*0}(2.6,0.7){$\sigma$}
		\rput[b]{*0}(1.5,0.0){$\psi$}
		\rput[b]{*0}(0.9,0.92){\scriptsize{$3$}}
		\rput[b]{*0}(2.1,0.92){\scriptsize{$3$}}
	\end{pspicture}
	}} \! \! \right),\label{eq:zs} \\
	\partial_{k}Z_{k}^{\,\pi} & = & \mathcal{V}^{-1} 
	\left.\frac{\mathrm{d}}{\mathrm{d}p^2}\right|_{p^{2}=0}
	\frac{\delta^{2}\partial_{k}\Gamma_{k}}{\delta\pi_{1}(-p)\delta\pi_{1}(p)} \nonumber\\
	& = & \mathcal{V}^{-1} \left.\frac{\mathrm{d}}{\mathrm{d}p^2}\right|_{p^{2}=0}\biggg(
	\frac{1}{2} \! \! \vcenter{\hbox{
	\begin{pspicture}(3.0,2.0)
		\psarc[linewidth=0.02,linestyle=dashed,dash=2pt 1pt,linecolor=blue](1.5,1.0){0.6}{20}{65}
		\psarc[linewidth=0.02,linestyle=dashed,dash=2pt 1pt,linecolor=blue](1.5,1.0){0.6}{115}{160}
		\psarc[linewidth=0.02,linestyle=dashed,dash=2pt 1pt,linecolor=red](1.5,1.0){0.6}{200}{340}
		\psline[linewidth=0.02,linestyle=dashed,dash=2pt 1pt,linecolor=red](0.1,1.0)(0.7,1.0)
		\psline[linewidth=0.02,linestyle=dashed,dash=2pt 1pt,linecolor=red](2.3,1.0)(2.9,1.0)
		\pscircle[linewidth=0.03,fillstyle=solid,fillcolor=NavyBlue](1.5,1.6){0.25}
		\psline[linewidth=0.03](1.5,1.6)(1.67,1.77)
		\psline[linewidth=0.03](1.5,1.6)(1.33,1.77)
		\psline[linewidth=0.03](1.5,1.6)(1.33,1.43)
		\psline[linewidth=0.03](1.5,1.6)(1.67,1.43)
		\pscircle[linewidth=0.03,fillstyle=solid,fillcolor=lightgray](0.9,1.0){0.20}
		\pscircle[linewidth=0.03,fillstyle=solid,fillcolor=lightgray](2.1,1.0){0.20}
		\rput[b]{*0}(2.1,1.4){$\sigma$}
		\rput[b]{*0}(0.9,1.4){$\sigma$}
		\rput[b]{*0}(0.4,0.7){$\pi$}
		\rput[b]{*0}(2.6,0.7){$\pi$}
		\rput[b]{*0}(1.5,0.15){$\pi$}
		\rput[b]{*0}(0.9,0.92){\scriptsize{$3$}}
		\rput[b]{*0}(2.1,0.92){\scriptsize{$3$}}
	\end{pspicture}
	}} \! \! \! 
	 + \frac{1}{2} \! \! \vcenter{\hbox{
	\begin{pspicture}(3.0,2.0)
		\psarc[linewidth=0.02,linestyle=dashed,dash=2pt 1pt,linecolor=red](1.5,1.0){0.6}{20}{65}
		\psarc[linewidth=0.02,linestyle=dashed,dash=2pt 1pt,linecolor=red](1.5,1.0){0.6}{115}{160}
		\psarc[linewidth=0.02,linestyle=dashed,dash=2pt 1pt,linecolor=blue](1.5,1.0){0.6}{200}{340}
		\psline[linewidth=0.02,linestyle=dashed,dash=2pt 1pt,linecolor=red](0.1,1.0)(0.7,1.0)
		\psline[linewidth=0.02,linestyle=dashed,dash=2pt 1pt,linecolor=red](2.3,1.0)(2.9,1.0)
		\pscircle[linewidth=0.03,fillstyle=solid,fillcolor=RedOrange](1.5,1.6){0.25}
		\psline[linewidth=0.03](1.5,1.6)(1.67,1.77)
		\psline[linewidth=0.03](1.5,1.6)(1.33,1.77)
		\psline[linewidth=0.03](1.5,1.6)(1.33,1.43)
		\psline[linewidth=0.03](1.5,1.6)(1.67,1.43)
		\pscircle[linewidth=0.03,fillstyle=solid,fillcolor=lightgray](0.9,1.0){0.20}
		\pscircle[linewidth=0.03,fillstyle=solid,fillcolor=lightgray](2.1,1.0){0.20}
		\rput[b]{*0}(2.1,1.4){$\pi$}
		\rput[b]{*0}(0.9,1.4){$\pi$}
		\rput[b]{*0}(0.4,0.7){$\pi$}
		\rput[b]{*0}(2.6,0.7){$\pi$}
		\rput[b]{*0}(1.5,0.15){$\sigma$}
		\rput[b]{*0}(0.9,0.92){\scriptsize{$3$}}
		\rput[b]{*0}(2.1,0.92){\scriptsize{$3$}}
	\end{pspicture}
	}} \nonumber\\
	& & \qquad \qquad \qquad \quad \ - \! \! \! \vcenter{\hbox{
	\begin{pspicture}(3.0,2.0)
		\psarc[linewidth=0.02,arrowsize=2pt 3,arrowinset=0]{->}(1.5,1.0){0.6}{20}{55}
		\psarc[linewidth=0.02](1.5,1.0){0.6}{46}{65}
		\psarc[linewidth=0.02,arrowsize=2pt 3,arrowinset=0]{->}(1.5,1.0){0.6}{115}{150}
		\psarc[linewidth=0.02](1.5,1.0){0.6}{141}{160}
		\psarc[linewidth=0.02,arrowsize=2pt 3,arrowinset=0]{->}(1.5,1.0){0.6}{200}{279}
		\psarc[linewidth=0.02](1.5,1.0){0.6}{270}{340}
		\psline[linewidth=0.02,linestyle=dashed,dash=2pt 1pt,linecolor=red](0.1,1.0)(0.7,1.0)
		\psline[linewidth=0.02,linestyle=dashed,dash=2pt 1pt,linecolor=red](2.3,1.0)(2.9,1.0)
		\pscircle[linewidth=0.03,fillstyle=solid,fillcolor=gray](1.5,1.6){0.25}
		\psline[linewidth=0.03](1.5,1.6)(1.67,1.77)
		\psline[linewidth=0.03](1.5,1.6)(1.33,1.77)
		\psline[linewidth=0.03](1.5,1.6)(1.33,1.43)
		\psline[linewidth=0.03](1.5,1.6)(1.67,1.43)
		\pscircle[linewidth=0.03,fillstyle=solid,fillcolor=lightgray](0.9,1.0){0.20}
		\pscircle[linewidth=0.03,fillstyle=solid,fillcolor=lightgray](2.1,1.0){0.20}
		\rput[b]{*0}(2.1,1.4){$\psi$}
		\rput[b]{*0}(0.9,1.4){$\psi$}
		\rput[b]{*0}(0.4,0.7){$\pi$}
		\rput[b]{*0}(2.6,0.7){$\pi$}
		\rput[b]{*0}(1.5,0.0){$\psi$}
		\rput[b]{*0}(0.9,0.92){\scriptsize{$3$}}
		\rput[b]{*0}(2.1,0.92){\scriptsize{$3$}}
	\end{pspicture}
	}} \! \! \! 
	- \frac{1}{2} \! \!
	\vcenter{\hbox{
	\begin{pspicture}[showgrid=false](2.0,2.0)
		\psarc[linewidth=0.02,linestyle=dashed,dash=2pt 1pt,linecolor=red](1.0,1.0){0.6}{290}{65}
		\psarc[linewidth=0.02,linestyle=dashed,dash=2pt 1pt,linecolor=red](1.0,1.0){0.6}{115}{250}
		\psline[linewidth=0.02,linestyle=dashed,dash=2pt 1pt,linecolor=red](0.35,0.1)(0.818408,0.316188)
		\psline[linewidth=0.02,linestyle=dashed,dash=2pt 1pt,linecolor=red](1.181592,0.316188)(1.65,0.1)
		\pscircle[linewidth=0.03,fillstyle=solid,fillcolor=RedOrange](1.0,1.6){0.25}
		\psline[linewidth=0.03](1.0,1.6)(1.17,1.77)
		\psline[linewidth=0.03](1.0,1.6)(0.83,1.77)
		\psline[linewidth=0.03](1.0,1.6)(0.83,1.43)
		\psline[linewidth=0.03](1.0,1.6)(1.17,1.43)
		\pscircle[linewidth=0.03,fillstyle=solid,fillcolor=lightgray](1.0,0.4){0.20}
		\rput[b]{*0}(0.985,0.32){\scriptsize{$4$}}
		\rput[b]{*0}(0.2,0.9){$\pi$}
		\rput[b]{*0}(1.8,0.9){$\pi$}
		\rput[b]{*0}(0.2,0.0){$\pi$}
		\rput[b]{*0}(1.8,0.0){$\pi$}
	\end{pspicture}
	}} \, \, \biggg),\label{eq:zp}  \\
	\partial_{k}Z_{k}^{\,\psi} & = & \frac{i}{4} \mathcal{V}^{-1}
	\left.\frac{\mathrm{d}}{\mathrm{d}p^2}\right|_{p^{2}=0}
	\tr_{\gamma} \left[\frac{\delta}{\delta \bar{\psi}(p)}\, \partial_{k}\Gamma_{k}\,
	\frac{\overleftarrow{\delta}}{\delta\psi (p)}\ \gamma_{\mu}p_{\mu} \right] \nonumber\\
	& & \frac{i}{4} \mathcal{V}^{-1}
	\left.\frac{\mathrm{d}}{\mathrm{d}p^2}\right|_{p^{2}=0}
	\tr_{\gamma} \biggg[\biggg(
	\frac{1}{2} \! \! \vcenter{\hbox{
	\begin{pspicture}(3.0,2.0)
		\psarc[linewidth=0.02,linestyle=dashed,dash=2pt 1pt,linecolor=blue](1.5,1.0){0.6}{20}{65}
		\psarc[linewidth=0.02,linestyle=dashed,dash=2pt 1pt,linecolor=blue](1.5,1.0){0.6}{115}{160}
		\psarc[linewidth=0.02,arrowsize=2pt 3,arrowinset=0]{->}(1.5,1.0){0.6}{200}{279}
		\psarc[linewidth=0.02](1.5,1.0){0.6}{270}{340}
		\psline[linewidth=0.02,ArrowInside=->,ArrowInsidePos=0.55,
		arrowsize=2pt 3,arrowinset=0](0.1,1.0)(0.7,1.0)
		\psline[linewidth=0.02,ArrowInside=->,ArrowInsidePos=0.55,
		arrowsize=2pt 3,arrowinset=0](2.3,1.0)(2.9,1.0)
		\pscircle[linewidth=0.03,fillstyle=solid,fillcolor=NavyBlue](1.5,1.6){0.25}
		\psline[linewidth=0.03](1.5,1.6)(1.67,1.77)
		\psline[linewidth=0.03](1.5,1.6)(1.33,1.77)
		\psline[linewidth=0.03](1.5,1.6)(1.33,1.43)
		\psline[linewidth=0.03](1.5,1.6)(1.67,1.43)
		\pscircle[linewidth=0.03,fillstyle=solid,fillcolor=lightgray](0.9,1.0){0.20}
		\pscircle[linewidth=0.03,fillstyle=solid,fillcolor=lightgray](2.1,1.0){0.20}
		\rput[b]{*0}(2.1,1.4){$\sigma$}
		\rput[b]{*0}(0.9,1.4){$\sigma$}
		\rput[b]{*0}(0.4,0.6){$\psi$}
		\rput[b]{*0}(2.6,0.6){$\psi$}
		\rput[b]{*0}(1.5,0.0){$\psi$}
		\rput[b]{*0}(0.9,0.92){\scriptsize{$3$}}
		\rput[b]{*0}(2.1,0.92){\scriptsize{$3$}}
	\end{pspicture}
	}} \! \! \! 
	+ \frac{1}{2} \! \! \vcenter{\hbox{
	\begin{pspicture}(3.0,2.0)
		\psarc[linewidth=0.02,linestyle=dashed,dash=2pt 1pt,linecolor=red](1.5,1.0){0.6}{20}{65}
		\psarc[linewidth=0.02,linestyle=dashed,dash=2pt 1pt,linecolor=red](1.5,1.0){0.6}{115}{160}
		\psarc[linewidth=0.02,arrowsize=2pt 3,arrowinset=0]{->}(1.5,1.0){0.6}{200}{279}
		\psarc[linewidth=0.02](1.5,1.0){0.6}{270}{340}
		\psline[linewidth=0.02,ArrowInside=->,ArrowInsidePos=0.55,
		arrowsize=2pt 3,arrowinset=0](0.1,1.0)(0.7,1.0)
		\psline[linewidth=0.02,ArrowInside=->,ArrowInsidePos=0.55,
		arrowsize=2pt 3,arrowinset=0](2.3,1.0)(2.9,1.0)
		\pscircle[linewidth=0.03,fillstyle=solid,fillcolor=RedOrange](1.5,1.6){0.25}
		\psline[linewidth=0.03](1.5,1.6)(1.67,1.77)
		\psline[linewidth=0.03](1.5,1.6)(1.33,1.77)
		\psline[linewidth=0.03](1.5,1.6)(1.33,1.43)
		\psline[linewidth=0.03](1.5,1.6)(1.67,1.43)
		\pscircle[linewidth=0.03,fillstyle=solid,fillcolor=lightgray](0.9,1.0){0.20}
		\pscircle[linewidth=0.03,fillstyle=solid,fillcolor=lightgray](2.1,1.0){0.20}
		\rput[b]{*0}(2.1,1.4){$\pi$}
		\rput[b]{*0}(0.9,1.4){$\pi$}
		\rput[b]{*0}(0.4,0.6){$\psi$}
		\rput[b]{*0}(2.6,0.6){$\psi$}
		\rput[b]{*0}(1.5,0.0){$\psi$}
		\rput[b]{*0}(0.9,0.92){\scriptsize{$3$}}
		\rput[b]{*0}(2.1,0.92){\scriptsize{$3$}}
	\end{pspicture}
	}} \! \! \! \nonumber\\
	& & \qquad \qquad \qquad \qquad \qquad \ - \! \! \vcenter{\hbox{
	\begin{pspicture}(3.0,2.0)
		\psarc[linewidth=0.02,arrowsize=2pt 3,arrowinset=0]{->}(1.5,1.0){0.6}{20}{55}
		\psarc[linewidth=0.02](1.5,1.0){0.6}{46}{65}
		\psarc[linewidth=0.02,arrowsize=2pt 3,arrowinset=0]{->}(1.5,1.0){0.6}{115}{150}
		\psarc[linewidth=0.02](1.5,1.0){0.6}{141}{160}
		\psarc[linewidth=0.02,linestyle=dashed,dash=2pt 1pt,linecolor=blue](1.5,1.0){0.6}{200}{340}
		\psline[linewidth=0.02,ArrowInside=->,ArrowInsidePos=0.55,
		arrowsize=2pt 3,arrowinset=0](0.7,1.0)(0.1,1.0)
		\psline[linewidth=0.02,ArrowInside=->,ArrowInsidePos=0.55,
		arrowsize=2pt 3,arrowinset=0](2.9,1.0)(2.3,1.0)
		\pscircle[linewidth=0.03,fillstyle=solid,fillcolor=gray](1.5,1.6){0.25}
		\psline[linewidth=0.03](1.5,1.6)(1.67,1.77)
		\psline[linewidth=0.03](1.5,1.6)(1.33,1.77)
		\psline[linewidth=0.03](1.5,1.6)(1.33,1.43)
		\psline[linewidth=0.03](1.5,1.6)(1.67,1.43)
		\pscircle[linewidth=0.03,fillstyle=solid,fillcolor=lightgray](0.9,1.0){0.20}
		\pscircle[linewidth=0.03,fillstyle=solid,fillcolor=lightgray](2.1,1.0){0.20}
		\rput[b]{*0}(2.1,1.4){$\psi$}
		\rput[b]{*0}(0.9,1.4){$\psi$}
		\rput[b]{*0}(0.4,0.6){$\psi$}
		\rput[b]{*0}(2.6,0.6){$\psi$}
		\rput[b]{*0}(1.5,0.15){$\sigma$}
		\rput[b]{*0}(0.9,0.92){\scriptsize{$3$}}
		\rput[b]{*0}(2.1,0.92){\scriptsize{$3$}}
	\end{pspicture}
	}} \! \! \! 
	- \! \! \! \vcenter{\hbox{
	\begin{pspicture}(3.0,2.0)
		\psarc[linewidth=0.02,arrowsize=2pt 3,arrowinset=0]{->}(1.5,1.0){0.6}{20}{55}
		\psarc[linewidth=0.02](1.5,1.0){0.6}{46}{65}
		\psarc[linewidth=0.02,arrowsize=2pt 3,arrowinset=0]{->}(1.5,1.0){0.6}{115}{150}
		\psarc[linewidth=0.02](1.5,1.0){0.6}{141}{160}
		\psarc[linewidth=0.02,linestyle=dashed,dash=2pt 1pt,linecolor=red](1.5,1.0){0.6}{200}{340}
		\psline[linewidth=0.02,ArrowInside=->,ArrowInsidePos=0.55,
		arrowsize=2pt 3,arrowinset=0](0.7,1.0)(0.1,1.0)
		\psline[linewidth=0.02,ArrowInside=->,ArrowInsidePos=0.55,
		arrowsize=2pt 3,arrowinset=0](2.9,1.0)(2.3,1.0)
		\pscircle[linewidth=0.03,fillstyle=solid,fillcolor=gray](1.5,1.6){0.25}
		\psline[linewidth=0.03](1.5,1.6)(1.67,1.77)
		\psline[linewidth=0.03](1.5,1.6)(1.33,1.77)
		\psline[linewidth=0.03](1.5,1.6)(1.33,1.43)
		\psline[linewidth=0.03](1.5,1.6)(1.67,1.43)
		\pscircle[linewidth=0.03,fillstyle=solid,fillcolor=lightgray](0.9,1.0){0.20}
		\pscircle[linewidth=0.03,fillstyle=solid,fillcolor=lightgray](2.1,1.0){0.20}
		\rput[b]{*0}(2.1,1.4){$\psi$}
		\rput[b]{*0}(0.9,1.4){$\psi$}
		\rput[b]{*0}(0.4,0.6){$\psi$}
		\rput[b]{*0}(2.6,0.6){$\psi$}
		\rput[b]{*0}(1.5,0.15){$\pi$}
		\rput[b]{*0}(0.9,0.92){\scriptsize{$3$}}
		\rput[b]{*0}(2.1,0.92){\scriptsize{$3$}}
	\end{pspicture}
	}} \! \! \biggg) \gamma_{\mu}p_{\mu} \biggg],\label{eq:zf}  \\
	\partial_{k}C_{2,k} & = & \frac{1}{2} \mathcal{V}^{-1} 
	\left.\frac{\mathrm{d}}{\mathrm{d}p^2}\right|_{p^{2}=0}
	\frac{\delta^{4}\partial_{k}\Gamma_{k}}{\delta\pi_{1}(p)
	\delta\pi_{2}(-p)\delta\pi_{1}(0)\delta\pi_{2}(0)} \nonumber\\
	& = & \frac{1}{2} \mathcal{V}^{-1} 
	\left.\frac{\mathrm{d}}{\mathrm{d}p^2}\right|_{p^{2}=0}\Biggg(
	- \frac{1}{2} \! \! \vcenter{\hbox{
	\begin{pspicture}[showgrid=false](3.0,2.0)
		\psarc[linewidth=0.02,linestyle=dashed,dash=2pt 1pt,linecolor=blue](1.5,1.0){0.6}{20}{65}
		\psarc[linewidth=0.02,linestyle=dashed,dash=2pt 1pt,linecolor=blue](1.5,1.0){0.6}{115}{160}
		\psarc[linewidth=0.02,linestyle=dashed,dash=2pt 1pt,linecolor=red](1.5,1.0){0.6}{290}{340}
		\psarc[linewidth=0.02,linestyle=dashed,dash=2pt 1pt,linecolor=red](1.5,1.0){0.6}{200}{250}
		\psline[linewidth=0.02,linestyle=dashed,dash=2pt 1pt,linecolor=red](0.85,0.1)(1.318408,0.316188)
		\psline[linewidth=0.02,linestyle=dashed,dash=2pt 1pt,linecolor=red](1.681592,0.316188)(2.15,0.1)
		\psline[linewidth=0.02,linestyle=dashed,dash=2pt 1pt,linecolor=red](0.1,1.0)(0.7,1.0)
		\psline[linewidth=0.02,linestyle=dashed,dash=2pt 1pt,linecolor=red](2.3,1.0)(2.9,1.0)
		\pscircle[linewidth=0.03,fillstyle=solid,fillcolor=NavyBlue](1.5,1.6){0.25}
		\psline[linewidth=0.03](1.5,1.6)(1.67,1.77)
		\psline[linewidth=0.03](1.5,1.6)(1.33,1.77)
		\psline[linewidth=0.03](1.5,1.6)(1.33,1.43)
		\psline[linewidth=0.03](1.5,1.6)(1.67,1.43)
		\pscircle[linewidth=0.03,fillstyle=solid,fillcolor=lightgray](1.5,0.4){0.20}
		\rput[b]{*0}(1.485,0.32){\scriptsize{$4$}}
		\pscircle[linewidth=0.03,fillstyle=solid,fillcolor=lightgray](0.9,1.0){0.20}
		\pscircle[linewidth=0.03,fillstyle=solid,fillcolor=lightgray](2.1,1.0){0.20}
		\rput[b]{*0}(0.9,0.92){\scriptsize{$3$}}
		\rput[b]{*0}(2.1,0.92){\scriptsize{$3$}}
		\rput[b]{*0}(0.4,0.7){$\pi$}
		\rput[b]{*0}(2.6,0.7){$\pi$}
		\rput[b]{*0}(0.7,0.0){$\pi$}
		\rput[b]{*0}(2.3,0.0){$\pi$}
		\rput[b]{*0}(2.1,1.4){$\sigma$}
		\rput[b]{*0}(0.9,1.4){$\sigma$}
		\rput[t]{*0}(2.1,0.55){$\pi$}
		\rput[t]{*0}(0.9,0.55){$\pi$}
	\end{pspicture}
	}} \! \! \! 
	- \frac{1}{2} \! \! \vcenter{\hbox{
	\begin{pspicture}(3.0,2.0)
		\psarc[linewidth=0.02,linestyle=dashed,dash=2pt 1pt,linecolor=red](1.5,1.0){0.6}{20}{65}
		\psarc[linewidth=0.02,linestyle=dashed,dash=2pt 1pt,linecolor=red](1.5,1.0){0.6}{115}{160}
		\psarc[linewidth=0.02,linestyle=dashed,dash=2pt 1pt,linecolor=blue](1.5,1.0){0.6}{290}{340}
		\psarc[linewidth=0.02,linestyle=dashed,dash=2pt 1pt,linecolor=blue](1.5,1.0){0.6}{200}{250}
		\psline[linewidth=0.02,linestyle=dashed,dash=2pt 1pt,linecolor=red](0.85,0.1)(1.318408,0.316188)
		\psline[linewidth=0.02,linestyle=dashed,dash=2pt 1pt,linecolor=red](1.681592,0.316188)(2.15,0.1)
		\psline[linewidth=0.02,linestyle=dashed,dash=2pt 1pt,linecolor=red](0.1,1.0)(0.7,1.0)
		\psline[linewidth=0.02,linestyle=dashed,dash=2pt 1pt,linecolor=red](2.3,1.0)(2.9,1.0)
		\pscircle[linewidth=0.03,fillstyle=solid,fillcolor=RedOrange](1.5,1.6){0.25}
		\psline[linewidth=0.03](1.5,1.6)(1.67,1.77)
		\psline[linewidth=0.03](1.5,1.6)(1.33,1.77)
		\psline[linewidth=0.03](1.5,1.6)(1.33,1.43)
		\psline[linewidth=0.03](1.5,1.6)(1.67,1.43)
		\pscircle[linewidth=0.03,fillstyle=solid,fillcolor=lightgray](1.5,0.4){0.20}
		\rput[b]{*0}(1.485,0.32){\scriptsize{$4$}}
		\pscircle[linewidth=0.03,fillstyle=solid,fillcolor=lightgray](0.9,1.0){0.20}
		\pscircle[linewidth=0.03,fillstyle=solid,fillcolor=lightgray](2.1,1.0){0.20}
		\rput[b]{*0}(0.9,0.92){\scriptsize{$3$}}
		\rput[b]{*0}(2.1,0.92){\scriptsize{$3$}}
		\rput[b]{*0}(0.4,0.7){$\pi$}
		\rput[b]{*0}(2.6,0.7){$\pi$}
		\rput[b]{*0}(0.7,0.0){$\pi$}
		\rput[b]{*0}(2.3,0.0){$\pi$}
		\rput[b]{*0}(2.1,1.4){$\pi$}
		\rput[b]{*0}(0.9,1.4){$\pi$}
		\rput[t]{*0}(2.1,0.55){$\sigma$}
		\rput[t]{*0}(0.9,0.55){$\sigma$}
	\end{pspicture}
	}} \vspace{-0.7cm} \nonumber\\
	& & \qquad \qquad \qquad \qquad \ + \frac{1}{2} \! \! \vcenter{\hbox{
	\begin{pspicture}[showgrid=false](2.6,2.0)
		\psarc[linewidth=0.02,linestyle=dashed,dash=2pt 1pt,linecolor=blue](1.3,1.0){0.6}{20}{65}
		\psarc[linewidth=0.02,linestyle=dashed,dash=2pt 1pt,linecolor=blue](1.3,1.0){0.6}{115}{160}
		\psarc[linewidth=0.02,linestyle=dashed,dash=2pt 1pt,linecolor=blue](1.3,1.0){0.6}{200}{340}
		\psline[linewidth=0.02,linestyle=dashed,dash=2pt 1pt,linecolor=red](1.98381,1.18159)(2.2,1.65)
		\psline[linewidth=0.02,linestyle=dashed,dash=2pt 1pt,linecolor=red](1.98381,0.818408)(2.2,0.35)
		\psline[linewidth=0.02,linestyle=dashed,dash=2pt 1pt,linecolor=red](0.616188,1.18159)(0.4,1.65)
		\psline[linewidth=0.02,linestyle=dashed,dash=2pt 1pt,linecolor=red](0.616188,0.818408)(0.4,0.35)
		\pscircle[linewidth=0.03,fillstyle=solid,fillcolor=NavyBlue](1.3,1.6){0.25}
		\psline[linewidth=0.03](1.3,1.6)(1.47,1.77)
		\psline[linewidth=0.03](1.3,1.6)(1.13,1.77)
		\psline[linewidth=0.03](1.3,1.6)(1.13,1.43)
		\psline[linewidth=0.03](1.3,1.6)(1.47,1.43)
		\pscircle[linewidth=0.03,fillstyle=solid,fillcolor=lightgray](0.7,1.0){0.20}
		\pscircle[linewidth=0.03,fillstyle=solid,fillcolor=lightgray](1.9,1.0){0.20}
		\rput[b]{*0}(0.685,0.92){\scriptsize{$4$}}
		\rput[b]{*0}(1.885,0.92){\scriptsize{$4$}}
		\rput[b]{*0}(0.2,0.2){$\pi$}
		\rput[b]{*0}(2.4,0.2){$\pi$}
		\rput[b]{*0}(0.2,1.7){$\pi$}
		\rput[b]{*0}(2.4,1.7){$\pi$}
		\rput[b]{*0}(1.9,1.4){$\sigma$}
		\rput[b]{*0}(0.7,1.4){$\sigma$}
		\rput[b]{*0}(1.3,0.15){$\sigma$}
	\end{pspicture}
	}} \! \! \!
	 + \frac{1}{2} \! \! \vcenter{\hbox{
	\begin{pspicture}(2.6,2.0)
		\psarc[linewidth=0.02,linestyle=dashed,dash=2pt 1pt,linecolor=red](1.3,1.0){0.6}{20}{65}
		\psarc[linewidth=0.02,linestyle=dashed,dash=2pt 1pt,linecolor=red](1.3,1.0){0.6}{115}{160}
		\psarc[linewidth=0.02,linestyle=dashed,dash=2pt 1pt,linecolor=red](1.3,1.0){0.6}{200}{340}
		\psline[linewidth=0.02,linestyle=dashed,dash=2pt 1pt,linecolor=red](1.98381,1.18159)(2.2,1.65)
		\psline[linewidth=0.02,linestyle=dashed,dash=2pt 1pt,linecolor=red](1.98381,0.818408)(2.2,0.35)
		\psline[linewidth=0.02,linestyle=dashed,dash=2pt 1pt,linecolor=red](0.616188,1.18159)(0.4,1.65)
		\psline[linewidth=0.02,linestyle=dashed,dash=2pt 1pt,linecolor=red](0.616188,0.818408)(0.4,0.35)
		\pscircle[linewidth=0.03,fillstyle=solid,fillcolor=RedOrange](1.3,1.6){0.25}
		\psline[linewidth=0.03](1.3,1.6)(1.47,1.77)
		\psline[linewidth=0.03](1.3,1.6)(1.13,1.77)
		\psline[linewidth=0.03](1.3,1.6)(1.13,1.43)
		\psline[linewidth=0.03](1.3,1.6)(1.47,1.43)
		\pscircle[linewidth=0.03,fillstyle=solid,fillcolor=lightgray](0.7,1.0){0.20}
		\pscircle[linewidth=0.03,fillstyle=solid,fillcolor=lightgray](1.9,1.0){0.20}
		\rput[b]{*0}(0.685,0.92){\scriptsize{$4$}}
		\rput[b]{*0}(1.885,0.92){\scriptsize{$4$}}
		\rput[b]{*0}(0.2,0.2){$\pi$}
		\rput[b]{*0}(2.4,0.2){$\pi$}
		\rput[b]{*0}(0.2,1.7){$\pi$}
		\rput[b]{*0}(2.4,1.7){$\pi$}
		\rput[b]{*0}(1.9,1.4){$\pi$}
		\rput[b]{*0}(0.7,1.4){$\pi$}
		\rput[b]{*0}(1.3,0.15){$\pi$}
	\end{pspicture}
	}} \! \! \!
	 - \frac{1}{2} \! \! \vcenter{\hbox{
	\begin{pspicture}[showgrid=false](3.0,3.2)
		\psarc[linewidth=0.02,linestyle=dashed,dash=2pt 1pt,linecolor=blue](1.5,1.6){0.6}{20}{65}
		\psarc[linewidth=0.02,linestyle=dashed,dash=2pt 1pt,linecolor=blue](1.5,1.6){0.6}{115}{160}
		\psarc[linewidth=0.02,linestyle=dashed,dash=2pt 1pt,linecolor=blue](1.5,1.6){0.6}{290}{340}
		\psarc[linewidth=0.02,linestyle=dashed,dash=2pt 1pt,linecolor=red](1.5,1.6){0.6}{200}{250}
		\psline[linewidth=0.02,linestyle=dashed,dash=2pt 1pt,linecolor=red](2.18381,1.78159)(2.4,2.25)
		\psline[linewidth=0.02,linestyle=dashed,dash=2pt 1pt,linecolor=red](2.18381,1.418408)(2.4,0.95)
		\psline[linewidth=0.02,linestyle=dashed,dash=2pt 1pt,linecolor=red](0.1,1.6)(0.7,1.6)
		\psline[linewidth=0.02,linestyle=dashed,dash=2pt 1pt,linecolor=red](1.5,0.2)(1.5,0.8)
		\pscircle[linewidth=0.03,fillstyle=solid,fillcolor=NavyBlue](1.5,2.2){0.25}
		\psline[linewidth=0.03](1.5,2.2)(1.67,2.37)
		\psline[linewidth=0.03](1.5,2.2)(1.33,2.37)
		\psline[linewidth=0.03](1.5,2.2)(1.33,2.03)
		\psline[linewidth=0.03](1.5,2.2)(1.67,2.03)
		\pscircle[linewidth=0.03,fillstyle=solid,fillcolor=lightgray](1.5,1.0){0.20}
		\rput[b]{*0}(1.5,0.92){\scriptsize{$3$}}
		\pscircle[linewidth=0.03,fillstyle=solid,fillcolor=lightgray](0.9,1.6){0.20}
		\pscircle[linewidth=0.03,fillstyle=solid,fillcolor=lightgray](2.1,1.6){0.20}
		\rput[b]{*0}(0.9,1.52){\scriptsize{$3$}}
		\rput[b]{*0}(2.085,1.52){\scriptsize{$4$}}
		\rput[b]{*0}(2.6,0.8){$\pi$}
		\rput[b]{*0}(2.6,2.3){$\pi$}
		\rput[b]{*0}(0.4,1.3){$\pi$}
		\rput[b]{*0}(1.3,0.4){$\pi$}
		\rput[b]{*0}(2.1,2.0){$\sigma$}
		\rput[b]{*0}(0.9,2.0){$\sigma$}
		\rput[t]{*0}(2.1,1.15){$\sigma$}
		\rput[t]{*0}(0.9,1.15){$\pi$}
	\end{pspicture} 
	}} \vspace{-0.6cm}  
	\nonumber\\
	& & \qquad \qquad \qquad \qquad \ - \frac{1}{2} \! \! \vcenter{\hbox{
	\begin{pspicture}(3.0,3.2)
		\psarc[linewidth=0.02,linestyle=dashed,dash=2pt 1pt,linecolor=red](1.5,1.6){0.6}{20}{65}
		\psarc[linewidth=0.02,linestyle=dashed,dash=2pt 1pt,linecolor=red](1.5,1.6){0.6}{115}{160}
		\psarc[linewidth=0.02,linestyle=dashed,dash=2pt 1pt,linecolor=red](1.5,1.6){0.6}{290}{340}
		\psarc[linewidth=0.02,linestyle=dashed,dash=2pt 1pt,linecolor=blue](1.5,1.6){0.6}{200}{250}
		\psline[linewidth=0.02,linestyle=dashed,dash=2pt 1pt,linecolor=red](2.18381,1.78159)(2.4,2.25)
		\psline[linewidth=0.02,linestyle=dashed,dash=2pt 1pt,linecolor=red](2.18381,1.418408)(2.4,0.95)
		\psline[linewidth=0.02,linestyle=dashed,dash=2pt 1pt,linecolor=red](0.1,1.6)(0.7,1.6)
		\psline[linewidth=0.02,linestyle=dashed,dash=2pt 1pt,linecolor=red](1.5,0.2)(1.5,0.8)
		\pscircle[linewidth=0.03,fillstyle=solid,fillcolor=RedOrange](1.5,2.2){0.25}
		\psline[linewidth=0.03](1.5,2.2)(1.67,2.37)
		\psline[linewidth=0.03](1.5,2.2)(1.33,2.37)
		\psline[linewidth=0.03](1.5,2.2)(1.33,2.03)
		\psline[linewidth=0.03](1.5,2.2)(1.67,2.03)
		\pscircle[linewidth=0.03,fillstyle=solid,fillcolor=lightgray](1.5,1.0){0.20}
		\rput[b]{*0}(1.5,0.92){\scriptsize{$3$}}
		\pscircle[linewidth=0.03,fillstyle=solid,fillcolor=lightgray](0.9,1.6){0.20}
		\pscircle[linewidth=0.03,fillstyle=solid,fillcolor=lightgray](2.1,1.6){0.20}
		\rput[b]{*0}(0.9,1.52){\scriptsize{$3$}}
		\rput[b]{*0}(2.085,1.52){\scriptsize{$4$}}
		\rput[b]{*0}(2.6,0.8){$\pi$}
		\rput[b]{*0}(2.6,2.3){$\pi$}
		\rput[b]{*0}(0.4,1.3){$\pi$}
		\rput[b]{*0}(1.3,0.4){$\pi$}
		\rput[b]{*0}(2.1,2.0){$\pi$}
		\rput[b]{*0}(0.9,2.0){$\pi$}
		\rput[t]{*0}(2.1,1.15){$\pi$}
		\rput[t]{*0}(0.9,1.15){$\sigma$}
	\end{pspicture}
	}} \! \! \! 
	- \frac{1}{2} \! \! \vcenter{\hbox{
	\begin{pspicture}(3.0,3.2)
		\psarc[linewidth=0.02,linestyle=dashed,dash=2pt 1pt,linecolor=blue](1.5,1.6){0.6}{20}{65}
		\psarc[linewidth=0.02,linestyle=dashed,dash=2pt 1pt,linecolor=blue](1.5,1.6){0.6}{115}{160}
		\psarc[linewidth=0.02,linestyle=dashed,dash=2pt 1pt,linecolor=red](1.5,1.6){0.6}{290}{340}
		\psarc[linewidth=0.02,linestyle=dashed,dash=2pt 1pt,linecolor=blue](1.5,1.6){0.6}{200}{250}
		\psline[linewidth=0.02,linestyle=dashed,dash=2pt 1pt,linecolor=red](0.816188,1.78159)(0.6,2.25)
		\psline[linewidth=0.02,linestyle=dashed,dash=2pt 1pt,linecolor=red](0.816188,1.41841)(0.6,0.95)
		\psline[linewidth=0.02,linestyle=dashed,dash=2pt 1pt,linecolor=red](2.3,1.6)(2.9,1.6)
		\psline[linewidth=0.02,linestyle=dashed,dash=2pt 1pt,linecolor=red](1.5,0.2)(1.5,0.8)
		\pscircle[linewidth=0.03,fillstyle=solid,fillcolor=NavyBlue](1.5,2.2){0.25}
		\psline[linewidth=0.03](1.5,2.2)(1.67,2.37)
		\psline[linewidth=0.03](1.5,2.2)(1.33,2.37)
		\psline[linewidth=0.03](1.5,2.2)(1.33,2.03)
		\psline[linewidth=0.03](1.5,2.2)(1.67,2.03)
		\pscircle[linewidth=0.03,fillstyle=solid,fillcolor=lightgray](1.5,1.0){0.20}
		\rput[b]{*0}(1.5,0.92){\scriptsize{$3$}}
		\pscircle[linewidth=0.03,fillstyle=solid,fillcolor=lightgray](0.9,1.6){0.20}
		\pscircle[linewidth=0.03,fillstyle=solid,fillcolor=lightgray](2.1,1.6){0.20}
		\rput[b]{*0}(0.885,1.52){\scriptsize{$4$}}
		\rput[b]{*0}(2.1,1.52){\scriptsize{$3$}}
		\rput[b]{*0}(0.4,0.8){$\pi$}
		\rput[b]{*0}(0.4,2.3){$\pi$}
		\rput[b]{*0}(2.6,1.3){$\pi$}
		\rput[b]{*0}(1.3,0.4){$\pi$}
		\rput[b]{*0}(2.1,2.0){$\sigma$}
		\rput[b]{*0}(0.9,2.0){$\sigma$}
		\rput[t]{*0}(2.1,1.15){$\pi$}
		\rput[t]{*0}(0.9,1.15){$\sigma$}
	\end{pspicture}
	}} \! \! \!
	- \frac{1}{2} \! \! \vcenter{\hbox{
	\begin{pspicture}(3.0,3.2)
		\psarc[linewidth=0.02,linestyle=dashed,dash=2pt 1pt,linecolor=red](1.5,1.6){0.6}{20}{65}
		\psarc[linewidth=0.02,linestyle=dashed,dash=2pt 1pt,linecolor=red](1.5,1.6){0.6}{115}{160}
		\psarc[linewidth=0.02,linestyle=dashed,dash=2pt 1pt,linecolor=blue](1.5,1.6){0.6}{290}{340}
		\psarc[linewidth=0.02,linestyle=dashed,dash=2pt 1pt,linecolor=red](1.5,1.6){0.6}{200}{250}
		\psline[linewidth=0.02,linestyle=dashed,dash=2pt 1pt,linecolor=red](0.816188,1.78159)(0.6,2.25)
		\psline[linewidth=0.02,linestyle=dashed,dash=2pt 1pt,linecolor=red](0.816188,1.41841)(0.6,0.95)
		\psline[linewidth=0.02,linestyle=dashed,dash=2pt 1pt,linecolor=red](2.3,1.6)(2.9,1.6)
		\psline[linewidth=0.02,linestyle=dashed,dash=2pt 1pt,linecolor=red](1.5,0.2)(1.5,0.8)
		\pscircle[linewidth=0.03,fillstyle=solid,fillcolor=RedOrange](1.5,2.2){0.25}
		\psline[linewidth=0.03](1.5,2.2)(1.67,2.37)
		\psline[linewidth=0.03](1.5,2.2)(1.33,2.37)
		\psline[linewidth=0.03](1.5,2.2)(1.33,2.03)
		\psline[linewidth=0.03](1.5,2.2)(1.67,2.03)
		\pscircle[linewidth=0.03,fillstyle=solid,fillcolor=lightgray](1.5,1.0){0.20}
		\rput[b]{*0}(1.5,0.92){\scriptsize{$3$}}
		\pscircle[linewidth=0.03,fillstyle=solid,fillcolor=lightgray](0.9,1.6){0.20}
		\pscircle[linewidth=0.03,fillstyle=solid,fillcolor=lightgray](2.1,1.6){0.20}
		\rput[b]{*0}(0.885,1.52){\scriptsize{$4$}}
		\rput[b]{*0}(2.1,1.52){\scriptsize{$3$}}
		\rput[b]{*0}(0.4,0.8){$\pi$}
		\rput[b]{*0}(0.4,2.3){$\pi$}
		\rput[b]{*0}(2.6,1.3){$\pi$}
		\rput[b]{*0}(1.3,0.4){$\pi$}
		\rput[b]{*0}(2.1,2.0){$\pi$}
		\rput[b]{*0}(0.9,2.0){$\pi$}
		\rput[t]{*0}(2.1,1.15){$\sigma$}
		\rput[t]{*0}(0.9,1.15){$\pi$}
	\end{pspicture}
	}} \nonumber\\
	& & \qquad \qquad \qquad \qquad \ + \frac{1}{2} \! \! \vcenter{\hbox{
	\begin{pspicture}[showgrid=false](3.0,3.2)
		\psarc[linewidth=0.02,linestyle=dashed,dash=2pt 1pt,linecolor=blue](1.5,1.6){0.8}{30}{71}
		\psarc[linewidth=0.02,linestyle=dashed,dash=2pt 1pt,linecolor=blue](1.5,1.6){0.8}{109}{150}
		\psarc[linewidth=0.02,linestyle=dashed,dash=2pt 1pt,linecolor=red](1.5,1.6){0.8}{180}{220}
		\psarc[linewidth=0.02,linestyle=dashed,dash=2pt 1pt,linecolor=blue](1.5,1.6){0.8}{250}{290}
		\psarc[linewidth=0.02,linestyle=dashed,dash=2pt 1pt,linecolor=red](1.5,1.6){0.8}{320}{0}
		\psline[linewidth=0.02,linestyle=dashed,dash=2pt 1pt,linecolor=red](2.35655,1.98865)(2.57274,2.45706)
		\psline[linewidth=0.02,linestyle=dashed,dash=2pt 1pt,linecolor=red](0.643447,1.98865)(0.427259,2.45706)
		\psline[linewidth=0.02,linestyle=dashed,dash=2pt 1pt,linecolor=red](0.957328,0.763086)(0.74114,0.294678)
		\psline[linewidth=0.02,linestyle=dashed,dash=2pt 1pt,linecolor=red](2.04267,0.763086)(2.25886,0.294678)
		\pscircle[linewidth=0.03,fillstyle=solid,fillcolor=NavyBlue](1.5,2.4){0.25}
		\psline[linewidth=0.03](1.5,2.4)(1.67,2.57)
		\psline[linewidth=0.03](1.5,2.4)(1.33,2.57)
		\psline[linewidth=0.03](1.5,2.4)(1.33,2.23)
		\psline[linewidth=0.03](1.5,2.4)(1.67,2.23)
		\pscircle[linewidth=0.03,fillstyle=solid,fillcolor=lightgray](2.27274,1.80706){0.20}
		\rput[b]{*0}(2.27274,1.72706){\scriptsize{$3$}}
		\pscircle[linewidth=0.03,fillstyle=solid,fillcolor=lightgray](0.727259,1.80706){0.20}
		\rput[b]{*0}(0.727259,1.72706){\scriptsize{$3$}}
		\pscircle[linewidth=0.03,fillstyle=solid,fillcolor=lightgray](1.04114,0.944678){0.20}
		\rput[b]{*0}(1.04114,0.864678){\scriptsize{$3$}}
		\pscircle[linewidth=0.03,fillstyle=solid,fillcolor=lightgray](1.95886,0.944678){0.20}
		\rput[b]{*0}(1.95886,0.864678){\scriptsize{$3$}}
		\rput[b]{*0}(2.77274,2.50706){$\pi$}
		\rput[b]{*0}(0.227259,2.50706){$\pi$}
		\rput[b]{*0}(0.54114,0.144678){$\pi$}
		\rput[b]{*0}(2.45886,0.144678){$\pi$}
		\rput[b]{*0}(2.1,2.3){$\sigma$}
		\rput[b]{*0}(0.9,2.3){$\sigma$}
		\rput[b]{*0}(1.5,0.55){$\sigma$}
		\rput[t]{*0}(2.45,1.35){$\pi$}
		\rput[t]{*0}(0.55,1.35){$\pi$}
	\end{pspicture}
	}} \! \! \! 	
	+ \frac{1}{2} \! \! \vcenter{\hbox{
	\begin{pspicture}(3.0,3.2)
		\psarc[linewidth=0.02,linestyle=dashed,dash=2pt 1pt,linecolor=red](1.5,1.6){0.8}{30}{71}
		\psarc[linewidth=0.02,linestyle=dashed,dash=2pt 1pt,linecolor=red](1.5,1.6){0.8}{109}{150}
		\psarc[linewidth=0.02,linestyle=dashed,dash=2pt 1pt,linecolor=blue](1.5,1.6){0.8}{180}{220}
		\psarc[linewidth=0.02,linestyle=dashed,dash=2pt 1pt,linecolor=red](1.5,1.6){0.8}{250}{290}
		\psarc[linewidth=0.02,linestyle=dashed,dash=2pt 1pt,linecolor=blue](1.5,1.6){0.8}{320}{0}
		\psline[linewidth=0.02,linestyle=dashed,dash=2pt 1pt,linecolor=red](2.35655,1.98865)(2.57274,2.45706)
		\psline[linewidth=0.02,linestyle=dashed,dash=2pt 1pt,linecolor=red](0.643447,1.98865)(0.427259,2.45706)
		\psline[linewidth=0.02,linestyle=dashed,dash=2pt 1pt,linecolor=red](0.957328,0.763086)(0.74114,0.294678)
		\psline[linewidth=0.02,linestyle=dashed,dash=2pt 1pt,linecolor=red](2.04267,0.763086)(2.25886,0.294678)
		\pscircle[linewidth=0.03,fillstyle=solid,fillcolor=RedOrange](1.5,2.4){0.25}
		\psline[linewidth=0.03](1.5,2.4)(1.67,2.57)
		\psline[linewidth=0.03](1.5,2.4)(1.33,2.57)
		\psline[linewidth=0.03](1.5,2.4)(1.33,2.23)
		\psline[linewidth=0.03](1.5,2.4)(1.67,2.23)
		\pscircle[linewidth=0.03,fillstyle=solid,fillcolor=lightgray](2.27274,1.80706){0.20}
		\rput[b]{*0}(2.27274,1.72706){\scriptsize{$3$}}
		\pscircle[linewidth=0.03,fillstyle=solid,fillcolor=lightgray](0.727259,1.80706){0.20}
		\rput[b]{*0}(0.727259,1.72706){\scriptsize{$3$}}
		\pscircle[linewidth=0.03,fillstyle=solid,fillcolor=lightgray](1.04114,0.944678){0.20}
		\rput[b]{*0}(1.04114,0.864678){\scriptsize{$3$}}
		\pscircle[linewidth=0.03,fillstyle=solid,fillcolor=lightgray](1.95886,0.944678){0.20}
		\rput[b]{*0}(1.95886,0.864678){\scriptsize{$3$}}
		\rput[b]{*0}(2.77274,2.50706){$\pi$}
		\rput[b]{*0}(0.227259,2.50706){$\pi$}
		\rput[b]{*0}(0.54114,0.144678){$\pi$}
		\rput[b]{*0}(2.45886,0.144678){$\pi$}
		\rput[b]{*0}(2.1,2.3){$\pi$}
		\rput[b]{*0}(0.9,2.3){$\pi$}
		\rput[b]{*0}(1.5,0.55){$\pi$}
		\rput[t]{*0}(2.45,1.35){$\sigma$}
		\rput[t]{*0}(0.55,1.35){$\sigma$}
	\end{pspicture}
	}} \! \! \! 	
	 - \! \! \! \vcenter{\hbox{
	\begin{pspicture}(3.0,3.2)
		\psarc[linewidth=0.02,arrowsize=2pt 3,arrowinset=0]{->}(1.5,1.6){0.8}{30}{60}
		\psarc[linewidth=0.02](1.5,1.6){0.8}{51}{71}
		\psarc[linewidth=0.02,arrowsize=2pt 3,arrowinset=0]{->}(1.5,1.6){0.8}{109}{139}
		\psarc[linewidth=0.02](1.5,1.6){0.8}{130}{150}
		\psarc[linewidth=0.02,arrowsize=2pt 3,arrowinset=0]{->}(1.5,1.6){0.8}{180}{208}
		\psarc[linewidth=0.02](1.5,1.6){0.8}{199}{220}
		\psarc[linewidth=0.02,arrowsize=2pt 3,arrowinset=0]{->}(1.5,1.6){0.8}{250}{277}
		\psarc[linewidth=0.02](1.5,1.6){0.8}{268}{290}
		\psarc[linewidth=0.02,arrowsize=2pt 3,arrowinset=0]{->}(1.5,1.6){0.8}{320}{348}
		\psarc[linewidth=0.02](1.5,1.6){0.8}{339}{0}
		\psline[linewidth=0.02,linestyle=dashed,dash=2pt 1pt,linecolor=red](2.35655,1.98865)(2.57274,2.45706)
		\psline[linewidth=0.02,linestyle=dashed,dash=2pt 1pt,linecolor=red](0.643447,1.98865)(0.427259,2.45706)
		\psline[linewidth=0.02,linestyle=dashed,dash=2pt 1pt,linecolor=red](0.957328,0.763086)(0.74114,0.294678)
		\psline[linewidth=0.02,linestyle=dashed,dash=2pt 1pt,linecolor=red](2.04267,0.763086)(2.25886,0.294678)
		\pscircle[linewidth=0.03,fillstyle=solid,fillcolor=gray](1.5,2.4){0.25}
		\psline[linewidth=0.03](1.5,2.4)(1.67,2.57)
		\psline[linewidth=0.03](1.5,2.4)(1.33,2.57)
		\psline[linewidth=0.03](1.5,2.4)(1.33,2.23)
		\psline[linewidth=0.03](1.5,2.4)(1.67,2.23)
		\pscircle[linewidth=0.03,fillstyle=solid,fillcolor=lightgray](2.27274,1.80706){0.20}
		\rput[b]{*0}(2.27274,1.72706){\scriptsize{$3$}}
		\pscircle[linewidth=0.03,fillstyle=solid,fillcolor=lightgray](0.727259,1.80706){0.20}
		\rput[b]{*0}(0.727259,1.72706){\scriptsize{$3$}}
		\pscircle[linewidth=0.03,fillstyle=solid,fillcolor=lightgray](1.04114,0.944678){0.20}
		\rput[b]{*0}(1.04114,0.864678){\scriptsize{$3$}}
		\pscircle[linewidth=0.03,fillstyle=solid,fillcolor=lightgray](1.95886,0.944678){0.20}
		\rput[b]{*0}(1.95886,0.864678){\scriptsize{$3$}}
		\rput[b]{*0}(2.77274,2.50706){$\pi$}
		\rput[b]{*0}(0.227259,2.50706){$\pi$}
		\rput[b]{*0}(0.54114,0.144678){$\pi$}
		\rput[b]{*0}(2.45886,0.144678){$\pi$}
		\rput[b]{*0}(2.1,2.3){$\psi$}
		\rput[b]{*0}(0.9,2.3){$\psi$}
		\rput[b]{*0}(1.5,0.4){$\psi$}
		\rput[t]{*0}(2.45,1.35){$\psi$}
		\rput[t]{*0}(0.55,1.35){$\psi$}
	\end{pspicture}
	}} \! \! \Biggg), \label{eq:c2} \\
	\partial_{k}Z_{2,k} & = & \frac{1}{4} \mathcal{V}^{-1} 
	\left.\frac{\mathrm{d}}{\mathrm{d}p^2}\right|_{p^{2}=0}
	\frac{\delta^{4}\partial_{k}\Gamma_{k}}{\delta\pi_{1}(p)
	\delta\pi_{2}(0)\delta\pi_{1}(-p)\delta\pi_{2}(0)} 
	 = \frac{1}{4} \mathcal{V}^{-1} 
	\left.\frac{\mathrm{d}}{\mathrm{d}p^2}\right|_{p^{2}=0}
	\bigg(\text{same diagrams as in Eq.\ (\ref{eq:c2})}\bigg)
	\label{eq:z2}, \\	
	\partial_{k}C_{3,k} & = & \frac{5}{576}\Bigg\lbrace 
	- \frac{208}{5}\,\partial_{k}C_{5,k} + 112\, \partial_{k}C_{6,k}
	- 32\, \partial_{k}C_{7,k} + \frac{224}{5}\, \partial_{k}C_{8,k}
	+ \mathcal{V}^{-1} \bigg[
	\frac{\partial}{\partial p_{1,\mu}}\frac{\partial}{\partial p_{2,\mu}}
	\frac{\partial}{\partial p_{3,\nu}}\frac{\partial}{\partial p_{1,\nu}} \nonumber\\
	& & \qquad\quad - \frac{7}{10}\frac{\partial}{\partial p_{1,\mu}}
	\frac{\partial}{\partial p_{2,\mu}}
	\frac{\partial}{\partial p_{3,\nu}}\frac{\partial}{\partial p_{2,\nu}}
	\bigg]\bigg|_{p_{1}=p_{2}=p_{3}=0}
	\frac{\delta^{4}\partial_{k}\Gamma_{k}}{\delta\pi_{1}(p_{1})
	\delta\pi_{2}(p_{2})\delta\pi_{1}(p_{3})\delta\pi_{2}(-p_{1}-p_{2}-p_{3})}
	\Bigg\rbrace , \label{eq:c3} \\
	\partial_{k}C_{4,k} & = & - \frac{2}{576}\Bigg\lbrace 
	16\, \partial_{k}C_{5,k} + 400\, \partial_{k}C_{6,k}
	- 32\, \partial_{k}C_{7,k} + 160\, \partial_{k}C_{8,k}
	+ \mathcal{V}^{-1} \bigg[
	\frac{\partial}{\partial p_{1,\mu}}\frac{\partial}{\partial p_{2,\mu}}
	\frac{\partial}{\partial p_{3,\nu}}\frac{\partial}{\partial p_{1,\nu}} \nonumber\\
	& & \qquad\quad\ - \frac{5}{2}\frac{\partial}{\partial p_{1,\mu}}
	\frac{\partial}{\partial p_{2,\mu}}
	\frac{\partial}{\partial p_{3,\nu}}\frac{\partial}{\partial p_{2,\nu}}
	\bigg]\bigg|_{p_{1}=p_{2}=p_{3}=0}
	\frac{\delta^{4}\partial_{k}\Gamma_{k}}{\delta\pi_{1}(p_{1})
	\delta\pi_{2}(p_{2})\delta\pi_{1}(p_{3})\delta\pi_{2}(-p_{1}-p_{2}-p_{3})}
	\Bigg\rbrace , \\
	\partial_{k}C_{5,k} & = & \frac{1}{96} \mathcal{V}^{-1} 
	\bigg[\frac{\partial}{\partial p_{2,\mu}}\frac{\partial}{\partial p_{2,\mu}}
	\frac{\partial}{\partial p_{2,\nu}}\frac{\partial}{\partial p_{3,\nu}} \nonumber\\
	& & \qquad\qquad\quad - \frac{1}{2}\frac{\partial}{\partial p_{2,\mu}}
	\frac{\partial}{\partial p_{2,\mu}}
	\frac{\partial}{\partial p_{2,\nu}}\frac{\partial}{\partial p_{2,\nu}}
	\bigg]\bigg|_{p_{2}=p_{3}=0}
	\frac{\delta^{4}\partial_{k}\Gamma_{k}}{\delta\pi_{1}(-p_{2}-p_{3})
	\delta\pi_{2}(p_{2})\delta\pi_{1}(p_{3})\delta\pi_{2}(0)}, \\
	\partial_{k}C_{6,k} & = & - \frac{1}{96}\Bigg\lbrace 
	- 160\, \partial_{k}C_{5,k} - 64\, \partial_{k}C_{7,k} + \mathcal{V}^{-1}
	\frac{\partial}{\partial p_{2,\mu}}\frac{\partial}{\partial p_{4,\mu}}
	\frac{\partial}{\partial p_{2,\nu}}\frac{\partial}{\partial p_{4,\nu}}
	\bigg|_{p_{2}=p_{4}=0}
	\frac{\delta^{4}\partial_{k}\Gamma_{k}}{\delta\pi_{1}(-p_{2}-p_{4})
	\delta\pi_{2}(p_{2})\delta\pi_{1}(0)\delta\pi_{2}(p_{4})} \nonumber\\
	& & \qquad\quad\, - \frac{1}{12} \mathcal{V}^{-1}
	\frac{\partial}{\partial p_{\mu}}
	\frac{\partial}{\partial p_{\mu}}
	\frac{\partial}{\partial p_{\nu}}\frac{\partial}{\partial p_{\nu}}
	\bigg|_{p=0}
	\frac{\delta^{4}\partial_{k}\Gamma_{k}}{\delta\pi_{1}(-p)
	\delta\pi_{2}(0)\delta\pi_{1}(p)\delta\pi_{2}(0)}
	\Bigg\rbrace , \\
	\partial_{k}C_{7,k} & = & - \frac{1}{384} \mathcal{V}^{-1}
	\frac{\partial}{\partial p_{\mu}}\frac{\partial}{\partial p_{\mu}}
	\frac{\partial}{\partial p_{\nu}}\frac{\partial}{\partial p_{\nu}}
	\bigg|_{p=0}
	\frac{\delta^{4}\partial_{k}\Gamma_{k}}{\delta\pi_{1}(-p)
	\delta\pi_{2}(p)\delta\pi_{1}(0)\delta\pi_{2}(0)}, \\
	\partial_{k}C_{8,k} & = & \frac{1}{96}\Bigg\lbrace 
	- 160\, \partial_{k}C_{5,k} - 64\, \partial_{k}C_{7,k} + \mathcal{V}^{-1}
	\frac{\partial}{\partial p_{2,\mu}}\frac{\partial}{\partial p_{4,\mu}}
	\frac{\partial}{\partial p_{2,\nu}}\frac{\partial}{\partial p_{4,\nu}}
	\bigg|_{p_{2}=p_{4}=0}
	\frac{\delta^{4}\partial_{k}\Gamma_{k}}{\delta\pi_{1}(-p_{2}-p_{4})
	\delta\pi_{2}(p_{2})\delta\pi_{1}(0)\delta\pi_{2}(p_{4})} \nonumber\\
	& & \qquad\ - \frac{5}{24} \mathcal{V}^{-1}
	\frac{\partial}{\partial p_{\mu}}
	\frac{\partial}{\partial p_{\mu}}
	\frac{\partial}{\partial p_{\nu}}\frac{\partial}{\partial p_{\nu}}
	\bigg|_{p=0}
	\frac{\delta^{4}\partial_{k}\Gamma_{k}}{\delta\pi_{1}(-p)
	\delta\pi_{2}(0)\delta\pi_{1}(p)\delta\pi_{2}(0)}
	\Bigg\rbrace \label{eq:c8} ,
\end{IEEEeqnarray}
\end{widetext}
where $p$ or $p_{i}$ with $i \in \lbrace 1,2,3,4 \rbrace$ denote the external momenta 
and the diagrams on the right-hand side represent the sum of all possible permutations of the 
external legs. From Eq.\ (\ref{eq:c3}) on we omitted the diagrammatic 
interpretation of the flow of the four-pion vertex. In the flow equations 
for $C_{3,k}$, $\ldots$ , $C_{8,k}$, the structure of the diagrams remains unchanged,
as we also indicated in Eq.\ (\ref{eq:z2}). Only the configuration of the 
external momenta is different.

The propagators of $\sigma$ and $\pi$ 
are highlighted in blue and red, respectively. This 
further applies to the regulator insertions each carrying a 
particular wave-function renormalization, cf. Eqs.\ (\ref{eq:reg1})
-- (\ref{eq:reg3}). The black fermion lines are provided with a 
specific direction.

Let us remark that the four-point vertices with two external 
$\sigma$ legs, which one would expect in the equation for 
$Z_{k}^{\,\sigma}$, do not contribute as they are $p$ independent. 
The same holds true for the four-point diagram with two external 
pion legs and a $\sigma$ looping around. This would be part of 
the flow equation for $Z_{k}^{\,\pi}$. Five- and six-point vertices 
on the right of Eqs.\ (\ref{eq:c2}) -- (\ref{eq:c8}) are
omitted.

As an alternative to the differential equations for the wave-function 
renormalization factors we could also have presented the anomalous 
dimensions $\eta_{k}^{\sigma}$, $\eta_{k}^{\pi}$, 
and $\eta_{k}^{\psi}$, since
\vspace{-10pt}
\begin{IEEEeqnarray}{rCl}
	\eta_{k}^{\sigma} & = & -k\, \partial_{k} \ln Z_{k}^{\,\sigma}, \\
	\eta_{k}^{\pi} & = & -k\, \partial_{k} \ln Z_{k}^{\,\pi}, \\
	\eta_{k}^{\psi} & = & -k\, \partial_{k} \ln Z_{k}^{\,\psi}.
\end{IEEEeqnarray}

We partly derived the flow equations by hand and additionally used the
\texttt{Mathematica} packages \texttt{FeynCalc} \cite{Mertig:1990an, 
Shtabovenko:2016sxi}, \texttt{DoFun} \cite{Huber:2011qr}, and 
\texttt{FormTracer} \cite{Cyrol:2016zqb}.

\bibliography{bib_LECs}

\end{document}